\documentclass[useAMS,usenatbib]{mn2e}
\usepackage{graphicx}
\usepackage{amssymb}
\usepackage{rotating}

\newcommand{\apj}{ApJ}
\newcommand{\apjl}{ApJL}
\newcommand{\aap}{A\&A}

\newcommand{\mnras}{MNRAS}

\newcommand{\apjs}{ApJS}
\newcommand{\MC}{\multicolumn}
\newcommand{\kms}{km~s$^{-1}$}

\newcommand{\HII}{H{\sc ii}}
\newcommand{\sunn}{$_{\odot}$}

\newcounter{qub}
\newcommand{\qq}{\addtocounter{qub}{1}\arabic{qub}}

\title[Nearby Void XMP dwarfs: candidate selection]
{XMP gas-rich dwarfs in Nearby Voids: candidate selection}
\author[S.A.~Pustilnik, E.S.~Egorova, Y.A.~Perepelitsyna, A.Y.~Kniazev]
{S.A.~Pustilnik,$^{1}$\thanks{E-mail: sap@sao.ru (SAP)} E.S.~Egorova,$^{4}$
Y.A.~Perepelitsyna,$^{1}$ A.Y.~Kniazev$^{2,3,4}$ \\
$^1$ Special Astrophysical Observatory of RAS, Nizhnij Arkhyz,
Karachai-Circassia 369167, Russia \\
$^2$ South Africa Astronomical Observatory, Cape Town, South Africa \\
$^3$ South African Large Telescope, Cape Town, South Africa  \\
$^4$ Sternberg Astronomical Institute of Moscow State University, Moscow, Russia \\
}

\begin{document}

\label{firstpage}

\date{Accepted December 1, 2019, Received August 14, 2019}

\pagerange{\pageref{firstpage}--\pageref{lastpage}} \pubyear{2019}

\maketitle

\begin{abstract} 

We introduce a project aimed  at systematically searching for eXtremely
Metal-Poor (XMP) very gas-rich blue dwarfs in voids in the nearby
Universe. Several such galaxies were first identified in the course
of an unbiased study of the galaxy population in the nearby Lynx-Cancer void.
These very rare and unusual galaxies appear to be the best proxies for
the so-called Very Young Galaxies (VYGs) defined recently in
the model simulations by \citet{Tweed18}. We discuss the main properties
of ten prototype objects residing in nearby voids and formulate
criteria to search for similar dwarfs in other voids.
The recently published sample of 1354 Nearby Void Galaxies (NVG) is used
to identify a subsample of 60 void dwarf XMP candidates. We provide a list of
these XMP candidates with their main parameters and finding
charts. These candidates are the subjects of subsequent spectral,
photometric and HI studies in the accompanying papers. Looking ahead,
with reference to the submitted accompanying papers, we find
that this study results in the discovery of many new XMP dwarfs
with 12+$\log$(O/H)$\sim$7.0--7.3 dex.
\end{abstract}


\section[]{Introduction}
\label{sec:intro}
\setcounter{figure}{0}

Since the discovery about a half century ago of the unusual
Blue Compact Dwarf Galaxy (BCDG) with extremely low gas
metallicity, IZw18 (=MRK~116) [Z$_{\rm gas} \sim$ Z\sunn/30,
\citet{Searle72}],
interest in these rare galaxies has suffered ups and downs.
One of the motivations to
search for other similar BCGs was a hope that such unusual dwarfs
could be used as proxies for primordial galaxies at high redshifts.
See e.g., the review by \citet{KO}.
Such objects are also the best probes of an important
cosmological parameter - the abundance of primordial Helium
\citep[e.g.][]{Izotov97,Aver12}.

Recent interest in such objects is caused
by the results of massive spectroscopic surveys of faint galaxies
mostly found in the SDSS \citep[][and references therein]{DR7,DR14}.
Several groups used this database to search for eXtremely Metal-Poor (XMP)
candidates and investigated them with high-quality follow-up
spectroscopy \citep[e.g.][]
{First_SDSS,Izotov06,IT07,ITG12,Izotov16,Sanchez16,Guseva17}.

Despite having spectral data for more than one million galaxies,
due to various observational selection effects and the survey limits,
gas metallicity at the level of Z $\lesssim$ Z\sunn/10,
or 12+$\log$(O/H) $\lesssim$ 7.69~dex  was only found in a few hundred
galaxies \citep{Izotov16,Izotov19DR14}.
To date, a dozen or fewer XMP galaxies have
a reliable value of Z$_{\rm gas} \lesssim$ Z\sunn/30
\citep[see][based on SDSS DR14 database]{Izotov19DR14}.

Meanwhile, other means were also used to search for such rare objects
during the last decade. The first one exploits the correlation of very low gas
metallicity with low galaxy luminosity and with elevated gas content.
A high sensitivity blind HI survey with the Arecibo radio telescope
[ALFALFA, \citet{ALFALFA}] over an area of $\sim$7000 sq. deg. resulted in
$\sim$35000 confident extragalactic HI-bearing objects with radial
velocities up to $\sim$16000~\kms. A substantial number of them appear
to be
rather close low-mass dwarfs with previously unknown radial velocities, mostly
due to the bias of the main optical redshift surveys.

The dedicated project, SHIELD \citep{Cannon11,McQuinn14,SHIELD},
aimed to study
a subsample of faint gas-rich ALFALFA dwarfs, and found several interesting
dwarfs, including the record-low object AGC198691 with 12+$\log$(O/H) = 7.02
\citep{Hirschauer16} at distance $\sim$11~Mpc.
One more XMP dwarf, Leo~P, with 12+$\log$(O/H) = 7.17, was found close
to the Local Group [D=1.65~Mpc, \citet{Skillman13}].

An 'almost' dark very LSB dwarf, Coma~P = AGC229386 at D =
5.5~Mpc, with a record high ratio MHI/$L_{\rm B}$ = 28
\citep{ComaP_HI, ComaP_HST}, could also fall into the same category
of objects. However, its estimated metallicity of stars of Z=0.002
\citep{ComaP_HST} that corresponds to 12+$\log$(O/H) = 7.8,
appears too high. So, it is difficult to assign this gas-rich dwarf to
the XMP category.

For this kind of candidate selection, the overall detection rate for new
XMP galaxies appears higher than for SDSS emission line galaxies.
However, the total number of XMP dwarfs found this way still remains very
small.

New selection methods of XMP candidates based on galaxy photometric properties
as extracted from the SDSS database have been suggested. First, galaxies
have been examined within the parametric space (colours, morphology, sizes)
typical
of Leo~P and similar XMP galaxies found among galaxies with the SDSS images.
In particular, \citet{James17} found 11 dwarfs with Z $\lesssim$ Z\sunn/10
among 50 blue diffuse dwarf (BDD) candidates. However, the lowest
values of O/H are not extremely low and all have 12+$\log$(O/H) $>$ 7.4~dex.

A similar method was used by \citet{LittleCub,Hsyu2018}. For their selected
candidate sample of $\sim$150 blue late-type galaxies, they found with
follow-up spectroscopy one good nearby XMP object (dubbed as 'Little Cub')
with 12+$\log$(O/H) = 7.13, while all their other
candidates have 12+$\log$(O/H) $>$ 7.4~dex.
Thus, summarizing, we conclude that due to various observational selection
effects, the search for new XMP galaxies is very difficult. Due to
the well established statistical relation between metallicity and luminosity,
one expects that the great majority of late-type galaxies with
$M_{\rm B,tot} > -13$ should have Z$_{\rm gas} \lesssim$ Z\sunn/10. However,
in reality such galaxies appear substantially more rare \citep{Sanchez17}.

Finally, we have suggested one more method for finding XMP dwarfs
\citep{PaperI,PaperVII}.
It is based on the confident finding of an overall metallicity deficiency
(on average, by a factor of $\sim$1.5) of late-type galaxies in
voids with respect to similar objects in denser environments (typical
groups) \citep{PaperVII,Eridanus}.
Moreover, it was found that $\sim$1/3 of about 20 of the faintest
Lynx-Cancer void late-type dwarfs show reduced gas O/H by a factor of 2--4
with respect to a control sample of similar galaxies in denser environments
\citep{PaperIV,PaperVII}.
 The same objects have the extremely high gas mass-fraction of 0.95--0.99 and
show blue colours in their outer regions.

If we extrapolate this finding based on rather small statistics to
late-type dwarfs in other nearby voids (so-called Nearby Void Galaxies (NVG)
sample, see Sect.~\ref{sec:NVG_sample}), we can expect roughly
1/3 of all $\sim$180 NVG objects with $M_{\rm B,tot} > -13$ to have
Z$_{\rm gas} \lesssim$~(Z\sunn/30--Z\sunn/20). While this is quite
a strong prediction, to check this, one needs good statistics, and hence,
high-quality spectroscopy of a major fraction of known intrinsically
faint NVG objects.

The important question regarding the minimal metallicity of gas in
galaxies in the nearby Universe is related to the history of
intergalactic medium (IGM)
enrichment. According to various indications, outside the sites of significant
galaxy concentration (massive galaxy groups and clusters), the local IGM is
considered largely to have a metallicity of 1--2 percent of the solar
value. The so-called 'cold accretion' of this gas to low-mass dark
matters halos is considered as one of the main processes of dwarf galaxy
build-up.
If the gas that forms stars in galaxies has been partly accreted
from the IGM through cold-flows, then the IGM metallicity sets a lower
limit for the metallicity in XMPs \citep[e.g.][and references therein]
{Sanchez14}.
It is worth mentioning that in the distant Universe,
at redshifts of $z \sim 3.1$ and $\sim$4.4, clear evidence is found for
much lower metallicity in the IGM, of $\sim$0.001 and even $\sim$0.0001
the solar value \citep{Cooke17,Robert19}.

For the local Universe galaxies, metallicity of $\sim$0.02 Z\sunn\ (or
12+$\log$(O/H) $\sim$ 7.0 dex) has been found for only a handful of objects.
Two of them, found recently by \citet{Izotov18,Izotov19a}, are actively
star-forming dwarfs (or blue compact dwarfs), at distances of $\sim$186
and 650 Mpc, respectively. As shown by \citet{Izotov18} with
the comprehensive SED fitting, in J0811+4730, the mass of the stellar
population with ages of 1--10 Gyr is smaller (or much smaller) than the
mass of the young population.

One of the caveats of their analysis is related to the small diameter
of used spectrograph fibers, $\sim$3\arcsec\ corresponding to the linear
diameter of $\sim$3~kpc at the galaxy distance. One cannot exclude the
possibility that outside the region of the current starburst,
which contributes the major part of the light in the SDSS spectrum of
this BCG, there exists a disc of old stars with
a mass comparable to or larger than the mass of the 'young' population.
With this reservation, these distant XMP dwarfs can be treated as
probable counterparts of the predicted
Very Young Galaxies (VYGs), that formed the majority of their stars
during the last one or so Gyr \citep{Tweed18}.

Moreover, as shown in the comprehensive analysis of all available data
for the prototype XMP blue compact galaxy IZw18 and its fainter companion
IZw18C by \citet{PO12},  despite the detection with HST of red giant
branch stars, the conclusion that the main stellar mass was formed
during
the very recent cosmological epoch ($\sim$1~Gyr) remains valid. The reddish
light at the periphery of the main star-forming regions is dominated by the
strong nebular emission of the ionized gas. The presence of this highly
inhomogeneous nebular emission in peripheral regions complicates
background
subtraction and may affect the analysis of the HST Color-Magnitude Diagram of
stars in these regions. For IZw18C, the analysis of all available data leads
to ages of stellar populations of $\lesssim$0.1~Gyr. Of course, one cannot
exclude the presence of faint emission of populations with cosmological age,
but their possible mass is lower than the mass of the visible
components.
That is, in this context, IZw18 and IZw18C also should be treated as the best
proxies for VYGs.

We mention also the results of the HST data analysis for the system
of IZw18 and IZw18C by \citet{Annibali2013}. They confirm the detection of the
substantial stellar population with ages of $\gtrsim$1~Gyr. Based on this
fact, the authors conclude that IZw18 is not a truly young galaxy forming
its first generation of stars. However, we use in this paper the relaxed
definition of Very Young Galaxy as suggested by \citet{Tweed18}. Within
this framework, with the current limits to their ancient star formation
history and the mass of stars with ages of a few to ten Gyrs, IZw18 and
IZw18C have properties consistent with their being VYGs.

The simulated VYGs appear both as central galaxies in the low-mass Dark Matter
halos and as their satellites. In the lowest mass range,
M$_{*} \lesssim 10^7$~M\sunn, the fraction of VYGs among satellites is
comparable to or larger than that for central galaxies \citep{Tweed18}.
It is interesting to compare this with current observational results.
In particular,
\citet{Sanchez17} in their observational study of very low metallicity dwarfs
conclude that they are the central galaxies. On the other hand,
among the known prototype XMP dwarfs in Table
\ref{tab:prop_summary} there are several fainter members of pairs or
triplets.

A few dwarfs with similar extremely low gas O/H [12+$\log$(O/H) $\lesssim$
7.0--7.1], but with much lower star formation rate (SFR), were found
in the nearby Lynx-Cancer void at distances of 10--25 Mpc
\citep{PaperVII,Hirschauer16}.
The majority of them also show blue colours in their outer parts, consistent
with a time since the onset of the main star formation (SF) episode of
$\lesssim$(1--3)~Gyr as well as
the extreme gas-mass fractions, M$_{\rm gas}$/M$_{\rm bary} \sim$0.96--0.99.
However, due to their low luminosity and faint emission lines,
they are usually bypassed by typical redshift surveys.

% Having in mind the mentioned above
After the discovery of the unusual void dwarfs which appeared to be
almost exclusively the least luminous blue LSB galaxies in
the Lynx-Cancer void, it is tempting to look for similar objects in other
nearby voids. This is now possible thanks to the recently published
'Nearby Void Galaxies' (NVG) sample \citep[][hereafter PTM19]{PTM19}.
See Sec.~\ref{sec:NVG_sample} for details. We use this sample to separate
a group of about sixty NVG low-luminosity late-type dwarfs as candidate
gas-rich XMP galaxies for follow-up spectroscopy with the SAO Big
Telescope Alt-azimuth \citep[BTA, 6-m telescope;][]{BTA}
and Southern African Large Telescope \citep[SALT;][]{Buck06,Dono06}
to determine their gas O/H.
Besides the follow-up spectroscopy, additional studies include optical
photometry and HI mapping.
All results of this study will be presented in a series of accompanying
papers, either submitted or in preparation.

The rest of this paper is arranged as follows. In Sec.~\ref{sec:NVG_sample}
we briefly summarize the Nearby Void Galaxies sample. In
Sec.~\ref{sec:XMP_prototype} we describe the small group of prototype XMP
void dwarfs  and their main properties. Sec.~\ref{sec:XMP_candidates}
presents the criteria which we applied to separate out the subsample
of candidate void XMP dwarfs together with the resulting list.
In Sec.~\ref{sec:discussion} we discuss the issues of XMP dwarfs and VYGs.
Finally, in Sec.~\ref{sec:conclusions} we summarize and present our
conclusions.

\section[]{Nearby Void Galaxies sample}
\label{sec:NVG_sample}

The 'Nearby Void Galaxies' catalogue is a collection of galaxies which fall
inside
25 voids in the volume with $R$ = 25 Mpc (PTM19). In total, 1354 galaxies
within these voids are catalogued.
The galaxies have absolute blue magnitudes $M_{\rm B}$ in the range of
--7.5 to --20.5, with the median $M_{\rm B,med}$ of $\sim$--15.2.
Of them, 1088 void galaxies reside in the 'inner' parts of void, having a
distance to the nearest bordering 'luminous' galaxy
$D_{\rm NN} \geq 2.0$ Mpc (with the median of $D_{\rm NN,med,inner}$=3.4~Mpc).
The rest of the NVG
sample objects reside in 'outer' parts of voids, closer to their borders.
Their median parameter $D_{\rm NN,med,outer}$ = 1.6~Mpc.

About 200 NVG sample objects fall within the Local Volume, that is the region
within $R$ = 11.0 Mpc. Thanks to their proximity, they are especially
promising for more detailed study, including the study of resolved
stellar populations with the HST and JWST.

The main NVG sample is important for several reasons related to the role
of environment in galaxy evolution.
This is, on the one hand, much deeper than any other void galaxy sample,
and, on the other hand, contains hundreds of low mass/luminosity dwarfs
which comprise the great majority of the void galaxy population.
This gives us an advantage in comparison to similar studies
based on more distant and more shallow void galaxy samples
\citep[e.g.,][]{Kreckel12,Kreckel15,Wegner19}
to look for the effect of void environments
on the most numerous low mass objects.

\section[]{Prototype unusual XMP gas-rich dwarfs}
\label{sec:XMP_prototype}

\begin{table*}
\caption{Parameters of known prototype void very gas-rich XMP dwarfs}
%\begin{tabular}{ccclccrcllll}
\begin{tabular}{cccccccccllp{2.5cm}}
\hline
IAU name     &  O/H          & $\frac{\rm MHI}{L_{\rm B}}$ &$\mu_{\rm B,0,c}$ & $M_{\rm B}$ &$(g-r)_0$  &$V_{\rm h}$& D    &logMHI    &logM$_{*}$ &$B_{\rm tot}$& \MC{1}{c}{Notes~~~~~~~~~~~~~~~~~~~} \\
	     & dex           &             &                       & mag         & mag     &             & Mpc  &         &          & mag        &       \\
 ~~~~1       & 2             & 3~~         & ~~4                   & 5           &  6      & 7           & 8    & 9~~~    & ~~10     & 11~~       & 12~~  \\
\hline
J0015+0104   & 7.03$\pm$.05$\dagger$&   2.4  &  24.5            & -14.07      &  0.19 &2031         & 28.4      &  8.32 &6.68 & 18.31     & in Eridanus void \\
J0113+0052   & 7.16$\pm$.05~~  &   4.2       &  ...             & -14.20      &  ...  &1175         & 16.2      &  8.52 &...  & 16.90     & UGC772, merg. \\ % in group NGC428 \\
J0706+3020   & 7.06$\pm$.05$\dagger$&  17.1  &  25:             & -12.47      &--0.08 & 952         & 16.9      &  8.41 &6.2: & 19.15     & in merg. tripl. UGC3672  \\
J0926+3343   & 7.12$\pm$.03~~  &   3.2       &  25.4            & -12.90      &  0.09 & 536         & 10.6      &  7.86 &6.11 & 17.34     &  \\
J0929+2502   & 7.10$\pm$.08$\dagger$&   2.4  &  24.1            & -12.95      &  0.16 &1661         & 25.7      &  7.75 &6.54 & 19.24     &       \\ % f_gas=0.95 and T_old ~3-6 Gyr
J0943+3326   & 7.02$\pm$.02~~  &   6.5       &  ...             & -10.47      &  0.05 & 514         & 11.0      &  7.19 &5.46 & 19.82     & AGC198691, pair with UGC5186  \\
J0956+2850   & 6.96-7.15~~     &   3.0       &  ...             & -13.9:      &  0.05 & 502         & 12.7      &  8.22:&6.6: & 16.7:     & DDO68B, in merg. tripl. \\
J2104--0035  & 7.07$\pm$.07~~  &   4.1       &  ...             & -13.20      &--0.04 &1395         & 17.2      &  8.29 &6.0: & 18.07     & PGC1139658, merg.?   \\
\hline
J0723+3622   & ...             &  11.9       &  24.4            & -11.76      &--0.01 & 970         & 16.0      &  7.98 &5.90 & 19.46     & in tripl. J0723+36 \\
J0723+3624   & ...             &  28.3       &  24.6            &  -9.56      &  0.08 & 938         & 16.0      &  7.46 &5.00 & 21.68     & in tripl. J0723+36 \\
%\hline
%J1232+2025  & ...             &  28.0       &  26.6            &  -9.75      &--0.05 & 1348        & 5.5       &  7.54 &5.63 & 19.06     & $^2$~Coma~P = AGC229385   \\ %
\hline
J0811+4730   & 6.98$\pm$.02~~  &   ...       &  ...             & -15.07g     & ...   &13323        & 182       &  ...  &6.24 & 21.37g    & Izotov et al. (2018)   \\
J1234+3901   & 7.03$\pm$.03~~  &   ...       &  ...             & -17.32g     & ...   &39863        & 546       &  ...  &7.13 & 21.92g    & Izotov et al. (2019a)   \\
\hline
\hline
\multicolumn{12}{p{16.2cm}}{Full description of columns is given in Sec.~\ref{sec:XMP_prototype}. We briefly recall their meaning. Col.~2: gas O/H in units 12+$\log$(O/H); } \\
\multicolumn{12}{p{16.2cm}}{Col.~3: MHI/$L_{\rm B}$  in solar units; Col.~4: corrected central surface brightness $\mu_{\rm B,0,c}$ in mag/$\Box{\arcsec}$; Col.~5: Absolute $B$-band} \\
\multicolumn{12}{p{16.2cm}}{magnitude. Superscript $g$ for the two distant XMPs relates to SDSS $g$-band. Col.~6: corrected for MW extinction total } \\
\multicolumn{12}{p{16.2cm}}{$g-r$ colour; Col.~7: heliocentric radial velocity in \kms; Col.~8: adopted distance in Mpc; Col.~9: $\log$ of galaxy HI-mass } \\
\multicolumn{12}{p{16.2cm}}{in M\sunn; Col.~10: $\log$ of stellar mass in M\sunn; Col.11: total $B$-band magnitude. Col. 12: Notes. We abbreviate triplet } \\
\multicolumn{12}{p{16.2cm}}{as tripl. (see sect.~\ref{ssec:scenarios} for detail) and merger/merging as merg. Low accuracy data are marked with (:).} \\
\multicolumn{12}{p{16.2cm}}{$\dagger$ O/H derived via the Strong Line method by Izotov et al. (2019b)  --0.03 dex, to match off-set from (O/H)$_{\rm Te}$. } \\
\end{tabular}
\label{tab:prop_summary}
\end{table*}

To increase the efficiency of XMP dwarf searches among the NVG
representatives,
we need, on the one hand, to better outline the observational peculiarities
of already known dwarfs in this group (based mainly on objects from the
Lynx-Cancer void).  On the other hand, the real range of parametric space of
void XMP galaxies could be wider than we know from their small already
known subgroup. To account for this fact, we may try to expand somewhat
the known ranges.

In  Table~\ref{tab:prop_summary} we summarize the properties of 8 currently
identified representatives of these unusual XMP dwarfs in the nearby voids and
two extremely gas-rich blue void dwarfs with unknown O/H, which very
likely belong to the same type.
Four of eight XMP gas-rich objects in the upper part of
Table~\ref{tab:prop_summary} (J0706+3020, J0926+3343, J0929+2502 and
J0956+2850) were found as a result of an unbiased
study of a sample of about a hundred galaxies in the nearby Lynx-Cancer
void \citep{DDO68,J0926,PaperIII,Triplet,PaperVI,PaperVII,U3672}. The entry
J0956+2850 corresponds to the smaller component of the well known dwarf
merging system DDO68 \citep{DDO68,Ekta2008,DDO68_HST} for which
12+$\log$(O/H) is shown to be below 7.15 dex \citep{DDO68,IT07,DDO68_OH}.
Since in such an advanced stage of merging it is difficult to recover
the pre-merger mass ratio, the fainter component parameters are adopted
with large uncertainties.

One more XMP dwarf in the Lynx-Cancer void, J0943+3326 (AGC198691 = Leoncino)
was found independently among the ALFALFA blind HI survey \citep{ALFALFA}
newly identified gas-rich objects \citep{Hirschauer16}.

In addition, three new XMP dwarfs were found via their emission-line spectra
in the SDSS database and later were identified as representatives in nearby
voids. In particular, J0015+0104, was found by \citet{Guseva09,Guseva17}
and was studied further as one of the Eridanus void dwarfs by
\citet{Pustilnik13} and \citet{Eridanus}.
Two more XMP dwarfs J0113+0052 and J2104--0035 were found in the SDSS by
\citet{Izotov06} and were further studied by \citet{Ekta2008,Moiseev2010}.

Two of the Lynx-Cancer void unusual dwarfs are members of the very gas-rich
triplet J0723+36. Their MHI/$L_{\rm B}$ ratios, $\sim$12 and $\sim$28,
respectively, fall among the highest known values of this empirical
parameter.
The related estimates of their stellar mass fractions
M$_{\rm star}$/M$_{\rm gas} <$ 0.01 \citep{Triplet} are also extremely
low. Both dwarfs have very low H$\alpha$ fluxes and their O/H ratios are hard
to obtain. However,
estimates of their possible evolutionary status allow one to assign them
to the same group of unusual void dwarfs \citep{Triplet}.
Since they have no O/H values, we put them in Table~\ref{tab:prop_summary}
below the main group, separated by a horizontal line.

We notice that all these prototype very gas-rich XMP dwarfs have blue
colours, namely $(g-r)_{\mathrm 0}$ $\lesssim$ 0.20, and mostly
$(g-r)_{\mathrm 0}$ $\lesssim$ 0.15. Such colour indices correspond to the
PEGASE \citep{pegase} stellar metallicity z=0.0004 instantaneous
evolutionary tracks with a time since the (localized) star formation
episode of $\lesssim$~1--2~Gyr or to the time of continuous SF with
constant SFR of $\lesssim 3-5$~Gyr, depending on the adopted initial mass
function and the $(u-g)$ color \citep{PaperIV}.

At the bottom of Table~\ref{tab:prop_summary}, separated from the main
table by a horizontal line, we also include two recently
discovered very distant unusual compact star-forming galaxies with the
record-low gas O/H ratio and very blue continuum with no traces of old stellar
populations, J0811+4730 and J1234+3901 \citep{Izotov18,Izotov19a}.
These dwarfs are probably the best candidates for
so-called Very Young Galaxies predicted in models \citep{Tweed18}.
Such XMP objects caught in the short phase of strong starburst
are very rare as already mentioned in Sec.~\ref{sec:intro}.
They are similar in their strong starbursts to the well known much
closer XMP BCGs IZw18 and SBS0335--052E.

The columns of Table~\ref{tab:prop_summary} are as follows:
Col.~1 -- brief IAU-type name; Col.~2 -- adopted parameter of gas
12+$\log$(O/H) with its 1~$\sigma$ error; Col.~3 --
HI-mass to blue luminosity ratio, in solar units, MHI/$L_{\rm B}$;
Col.~4 -- $\mu_{\rm B,0,c}$ - (when known) central surface
brightness, corrected for extinction in the Milky Way and inclination
\citep{PaperIV};
Col.~5 -- absolute blue magnitudes $M_{\rm B}$, derived from $B_{\rm tot}$
in Col.~10 and Distance in Col.~8;
Col.~6 -- integrated colour $(g-r)_{\mathrm 0}$. This comes for
J0015+0104 from \citet{Pustilnik13}, for J0706+3020 from \citet{U3672},
for J0926+3343, J0929+2502, J0723+3622 and J0723+3624 from \citet{PaperIV},
for J0943+3326 we adopt $(g-r)$ presented in the SDSS DR14 database
\citep{DR14}.
For J0956+2850 (DDO68B), the range of colours for various parts of the assumed
fainter component of the merger comes from \citet{DDO68sdss}, for J2104--0049
we adopt our unpublished measurement from the SDSS images;
Col.~7 -- $V_{\rm h}$ -heliocentric velocity; Col.8 -- distance to object,
in Mpc, either known from TRGB estimate, or for the radial velocity based
distance - accounting for the large peculiar velocity of the Local Sheet,
up to $\sim$350~\kms, depending on the sky position
\citep{Tully08}. See PTM19 for more detail;
Col.~9 --   mass of HI-gas, in units of M\sunn;
Col.~10 --  estimate of stellar mass following the recipe by
\citet{Zibetti09}, in units of M\sunn. These estimates come
from \citet{Pustilnik13,PaperIV,U3672};
Col.11 --  total apparent $B$-band magnitude;
Col.12 --  Notes  with additional information and references
for parameters in the Table. Several entries in the Table with lower
accuracy are marked with a colon (:).

Summarizing the data in this table and in accord with our earlier
conclusions on the group properties of these unusual void dwarfs with the
lowest O/H ratios, we emphasize their low luminosity, $M_{\rm B}$
between $\sim$ --9.5 and --14.2, and medium to high MHI/$L_{\rm B}$,
from $\sim$2 up to 28. In most cases they are related to LSB discs
with blue colours.
The full range of baryonic mass (gas plus stars)  in known prototype
XMP void dwarfs is ($\sim$2--40) $\times$ 10$^7$~M\sunn. The atomic gas
HI along with atomic He (adopted as 30\% of HI) is dominant by mass
over the stellar component by $\sim$1.5--2 orders of magnitude.
Their dynamical masses M$_{\rm dyn}$
within the HI-radius at the surface density level of 0.3~M\sunn~pc$^{-2}$
are estimated only for several objects and typically comprise 5--10
M$_{\rm bary}$. The total masses of the related Cold Dark Matter (CDM)
halos, M$_{\rm vir}$, can be several to ten times higher than M$_{\rm dyn}$
as argued in simulations by \citet{Hoeft06}. That is, the expected range of
CDM halo masses for these XMP dwarfs is between $<10^9$ and
$\sim 10^{10}$ M\sunn.

\section[]{Candidate selection criteria and the list}
\label{sec:XMP_candidates}

We proceed with the selection of void XMP candidates based on the properties
of known void XMP dwarfs summarized in the previous section.
For this, we make the selection in several steps.

First, we use the NVG sample (PTM19) of 1354 galaxies residing in the Nearby
Voids, briefly described in Sec.~\ref{sec:NVG_sample}, to select objects
in its low luminosity tail. This step is related to our previous results as
explained below.

In the unbiased photometric and spectroscopic study of about a hundred
galaxies residing in a nearby Lynx-Cancer void \citep{PaperIV,PaperVII},
we found that the most gas-rich dwarfs, with the lowest gas O/H ratio and
{\it atypically} blue colours of stars at their periphery, are predominantly
the {\it lowest} luminosity dwarfs. Their fraction among
void galaxies is significantly elevated for the range of $M_{\rm B}> -13.2$.
However, inclusion in Table~\ref{tab:prop_summary} of three prototype XMP
dwarfs residing in other nearby voids, pushes this limit to $M_{\rm B}$
of --14.2. This indicates that the latter cut can be more appropriate for
picking up similar void XMP candidates.
Thus, allowing a small margin in this cut, we adopt a $B$-band absolute
magnitude limit of $-14.3$. This gives us an initial subsample of
$\sim$380 of the NVG faint-end galaxies.

Second, we use publicly available parameters of this subsample to further
select objects with substantially high value of MHI/$L_{\rm B} \gtrsim$1,
in particular as obtained from HyperLEDA
\citep{LEDA}\footnote{http://leda.univ-lyon1.fr} and NED.
At the first pass we obtained several tens galaxies with
MHI/$L_{\rm B} \gtrsim$ 3. Galaxies with that high a MHI/$L_{\rm B}$
ratio are quite rare. To be confident that this is not due to some errors,
we checked for alternative total blue
magnitudes $B_{\rm tot}$ in the literature, where possible.
Also, we used the SDSS image database.
We obtained their total magnitudes in circular apertures
in $g,r$ filters and used the transform of \citet{Lupton05} to derive their
$B_{\rm tot}$.

In more than half of such 'gas-rich' galaxies selected automatically, we found
that the updated $B_{\rm tot}$, taken either as a more reliable estimate
in the literature, or estimated by us from the SDSS images,
appear significantly brighter than
given in HyperLEDA. For some of the checked objects the value of HI-flux
in HyperLEDA was outdated and substantially larger than the more recent
measurements. All these updates were taken into account for the final
selection and the false 'gas-rich' galaxies were excluded. However, some of
the errors in $B_{\rm tot}$ and HI-flux for the selected XMP candidate
objects were detected later, when a galaxy was already observed in our
program. Their MHI/$L_{\rm B}$ ratio seems to be smaller than 1. We
kept, however, several such candidates in the list.
Therefore, parameters of $B_{\rm tot}$ and $M_{\rm B}$ shown
for the selected XMP candidates in Table~\ref{tab:list_candidate} can differ
significantly from those in the original NVG catalogue, which was based mostly
on data from HyperLEDA.

Third, where possible, we also determined SDSS integrated colours for the
selected galaxies. We use colour $g-r$, corrected for the MW reddening, as an
additional selection criterion, to cut-off the reddest objects. The goal
of this additional criterion is to pick up the objects with a predominant
contribution of light from a relatively young stellar population.
However, we chose to extend significantly the colour range defined
XMP prototype group. On the one hand, we want to account for possible
uncertainties in the XMP candidate colour estimates.
On the other hand, thanks to a wider range of $(g-r)$, we can examine
how tightly the very low gas metallicity is related to galaxy colour.
Therefore we adopt as a colour selection criterion $(g-r)_{0} < 0.4$.
We also kept in the
list several galaxies with somewhat doubtful $(g-r)_{0} > 0.4$ and large
MHI/$L_{\rm B}$.

As a result, from the original NVG subsample of $\sim$380 faint void
galaxies, applying additional selection criteria, we compiled
a subsample of sixty void dwarfs for further optical spectroscopy
(see Table~\ref{tab:list_candidate}).

For several of these candidates, SDSS spectra of the central part of
the galaxy
are available. Due to the redshift range of our sample, the important line
[O{\sc ii}]$\lambda$3727 used in the classical $T_{\rm e}$ method to estimate
O/H in \HII-regions, is outside the SDSS spectral coverage.
However, the other strong lines can provide useful information on the
expected level of the oxygen abundance. In particular, we exploit
the equations (1) and (3) from \citet{Izotov19DR14} to look for potential XMP
candidates. According to their equation (1), galaxies with parameter
R$_{\mathrm 23} \lesssim$4 should have 12+$\log$(O/H) $\lesssim$7.4. Here,
R$_{\mathrm 23}$=[I([O{\sc iii}]4959)+I([O{\sc iii}]5007)+I([O{\sc ii}]3727)]/I(H$\beta$).
I([O{\sc ii}]3727) is fitted with their equation (3), namely,
  I([O{\sc ii}]3727) = 20$\times$I([N{\sc ii}]6584),
when I([N{\sc ii}]6584)/I(H$\beta$)$<$0.1.

This information was also
taken into account in our selection of void XMP candidates to include
several objects with the value of MHI/$L_{\rm B} < 1$ or without HI-data,
 or with $g-r$ somewhat redder than $\sim$0.4.
Among XMP candidates in Table~\ref{tab:list_candidate} with SDSS spectra,
an estimate of O/H is available only for one, J1522+4201
\citep{Izotov19DR14}.

Besides, one more galaxy with published O/H, KK246, was included by us
as an XMP void candidate according to our selection criteria. Its value of
12+$\log$(O/H)=8.2 \citep{KK246_OH} appears a strong upward outlier with
respect of the value expected for its properties. Therefore it deserves an
independent check.

For several candidates in Table~\ref{tab:list_candidate}, after our additional
checks, the parameter MHI/$L_{\rm B}$ appeared substantially lower
than the one adopted initially. However, since their spectra were already
obtained, we kept them in the list. This allows us to address the diversity
of properties of void XMP dwarfs.

The selected candidates for spectral observations are listed
in Table~\ref{tab:list_candidate}. The columns include the following
information: Col.~1 -- {\it id} number; Col.~2 -- galaxy name
in the NVG catalogue, mainly adopted from the HyperLEDA database;
Col.~3  -- J2000 epoch coordinates; Col.~4 -- heliocentric velocity in
\kms; Col.~5 -- Distance in Mpc (either measured independently of velocity,
or with the use of the peculiar velocity correction similar to that described
for Table~\ref{tab:prop_summary}); Col.~6 -- estimate of total $B$-band
magnitude; Col.~7 -- the related absolute magnitude $M_{\rm B}$;
Col.~8 -- HI mass to the blue luminosity ratio MHI/$L_{\rm B}$,
in solar units; Col.~9 -- $\log$MHI and Col.~10 -- $\log$(M$_{*}$),
both in solar masses; Col.~11 -- Notes.
Where possible, we measured [similar to \citet{PaperIV}]
integrated colours of the examined candidates. Most
estimates of M$_{*}$ are obtained
via the method of \citet{Zibetti09}, based on our photometry of the
candidate galaxies available in the SDSS image database \citep{DR14}. For
several galaxies outside the SDSS footprint, where possible, we
performed similar photometry on the public databases of PanSTARRS PS1
\citep{PS1-database} and Dark Energy Survey (DES DR1) \citep{DESDR1}.
For several galaxies not in these databases we used estimates of
M$_{*}$ from \citet{Gurovich2010} for HIPASSJ0653-73,
\citet{Kirbi2008,Kreckel2011} for KK246 and \citet{IKar2014} for ESO238-005.

As one can see, XMP void candidates in Table~\ref{tab:list_candidate} fall
in a wide range of distances from $\sim$5 to 25~Mpc. Their $B_{\rm tot}$
range between $\sim$15.5 and $\sim$20.0~mag. Their absolute blue magnitudes
$M_{\rm B}$ fall between $-10.3$ and $-14.3$~mag. The ratio MHI/$L_{\rm B}$
for the great majority of selected candidates is between $\sim$1 and 6.5.
However,
for a couple of objects it was estimated as $\sim$0.6. The baryonic mass range
of the candidates is close to that of XMP prototypes, that is
(2--40)$\times$10$^7$~M\sunn. Only four objects have this parameter in the
range (0.6--2)$\times$10$^7$~M\sunn. As follows from comparison of mass
estimates of HI and stars in columns 9 and 10, for 48 galaxies which
have both estimates, the ratio M(HI+He)/M$_{*}$
for the candidate galaxies varies over a wide range, with extremes of
$\sim$2 and $\sim$100, with the median of 12, and with half of
the entries between 5 and 26. The future
analysis of measured gas O/H in these XMP candidates along with their
parameters
in Table~\ref{tab:list_candidate} will allow us to improve the current
selection criteria and to make the spectral follow-ups more efficient.

\begin{figure}
\includegraphics[width=6.5cm,angle=-90,clip=]{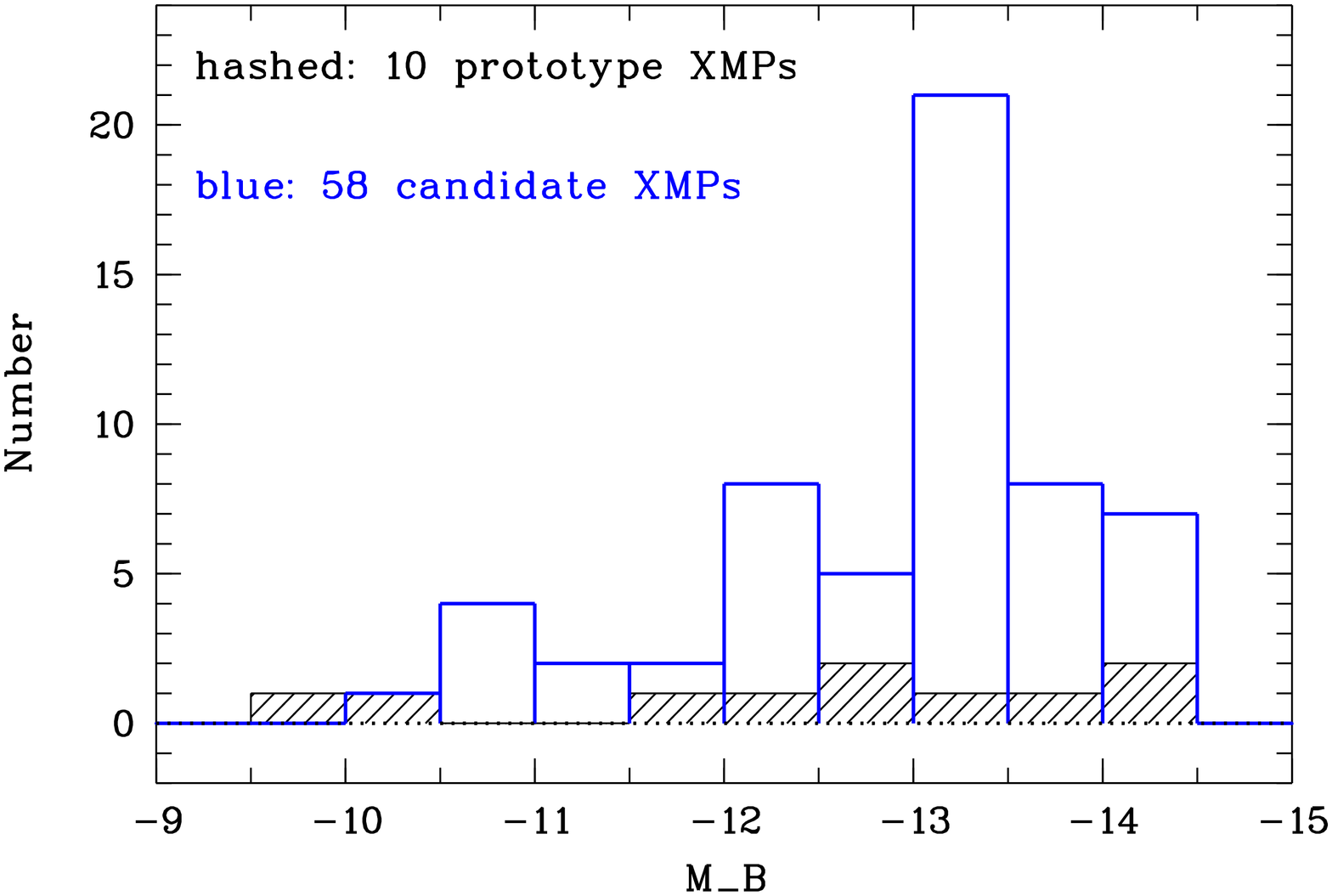}
\includegraphics[width=6.5cm,angle=-90,clip=]{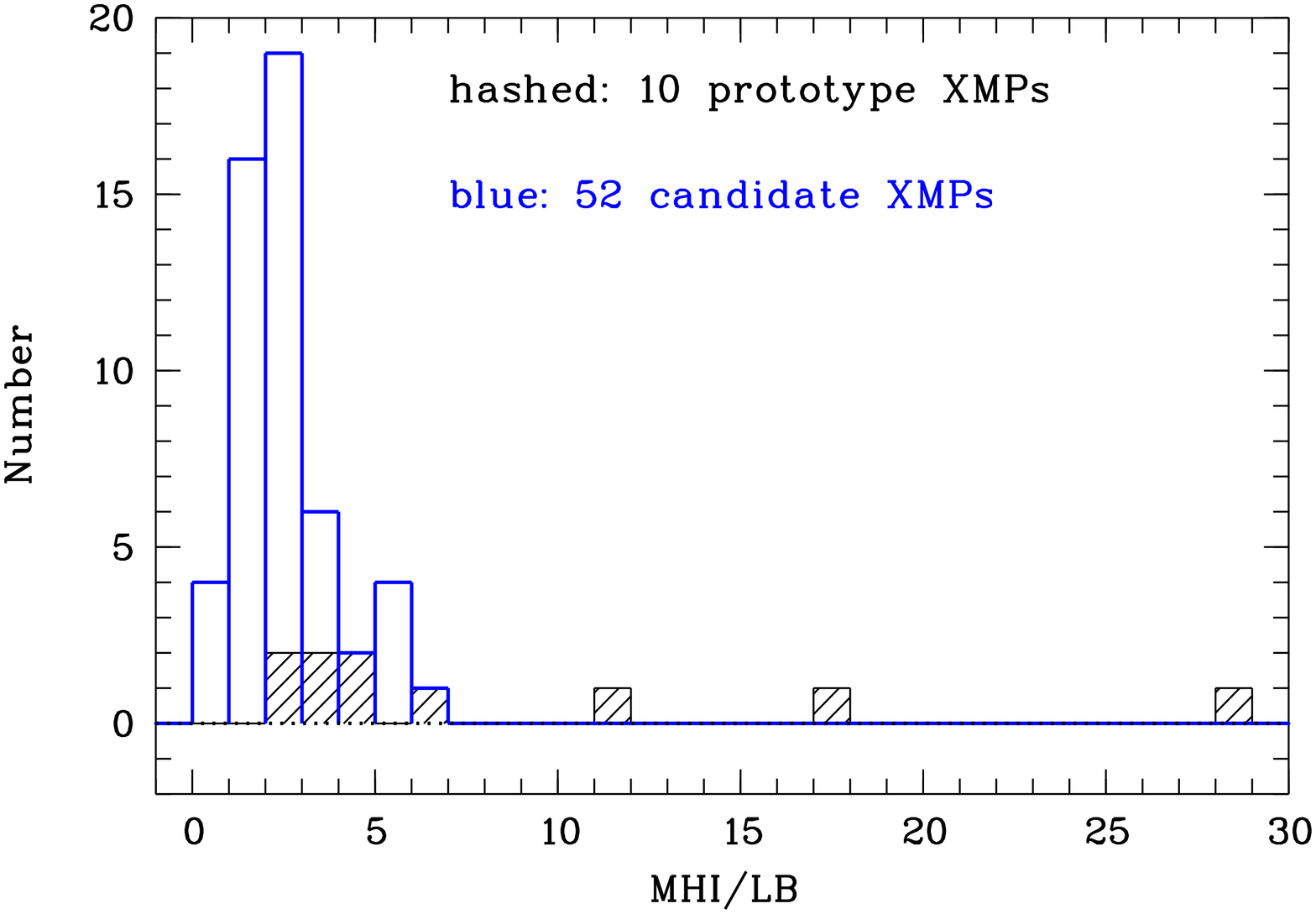}
\caption{Comparison of observational properties of 10 prototype XMP void
galaxies (hashed histograms) and XMP void candidates with available data
(blue thick histograms).
{\bf Top panel:} Distribution of absolute blue magnitudes $M_{\rm B}$.
{\bf Bottom panel:} Distribution of ratio MHI/$L_{\rm B}$ (in solar units). }
\label{fig:histo1}
\end{figure}

\begin{figure}
\includegraphics[width=6.5cm,angle=-90,clip=]{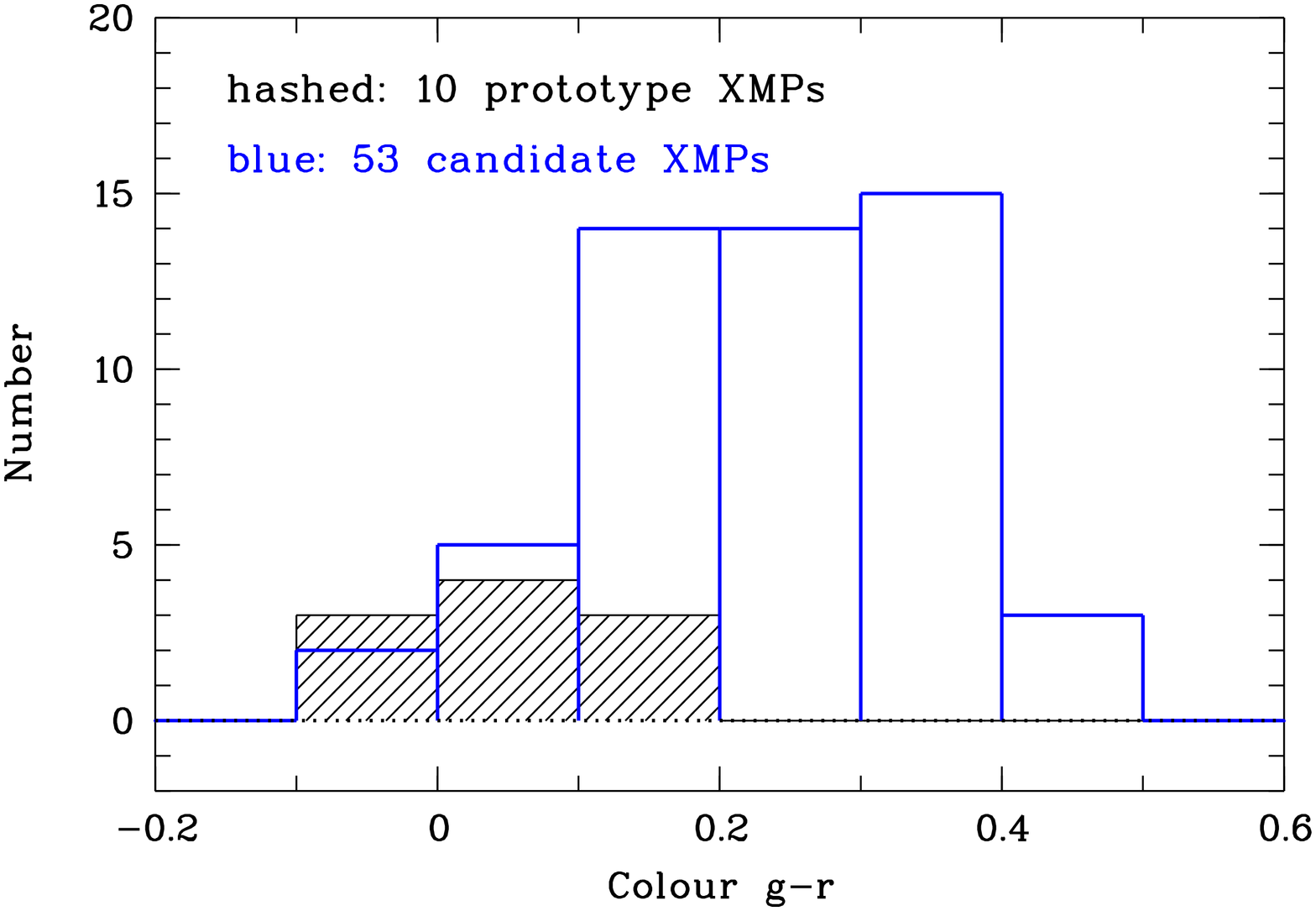}
\includegraphics[width=6.5cm,angle=-90,clip=]{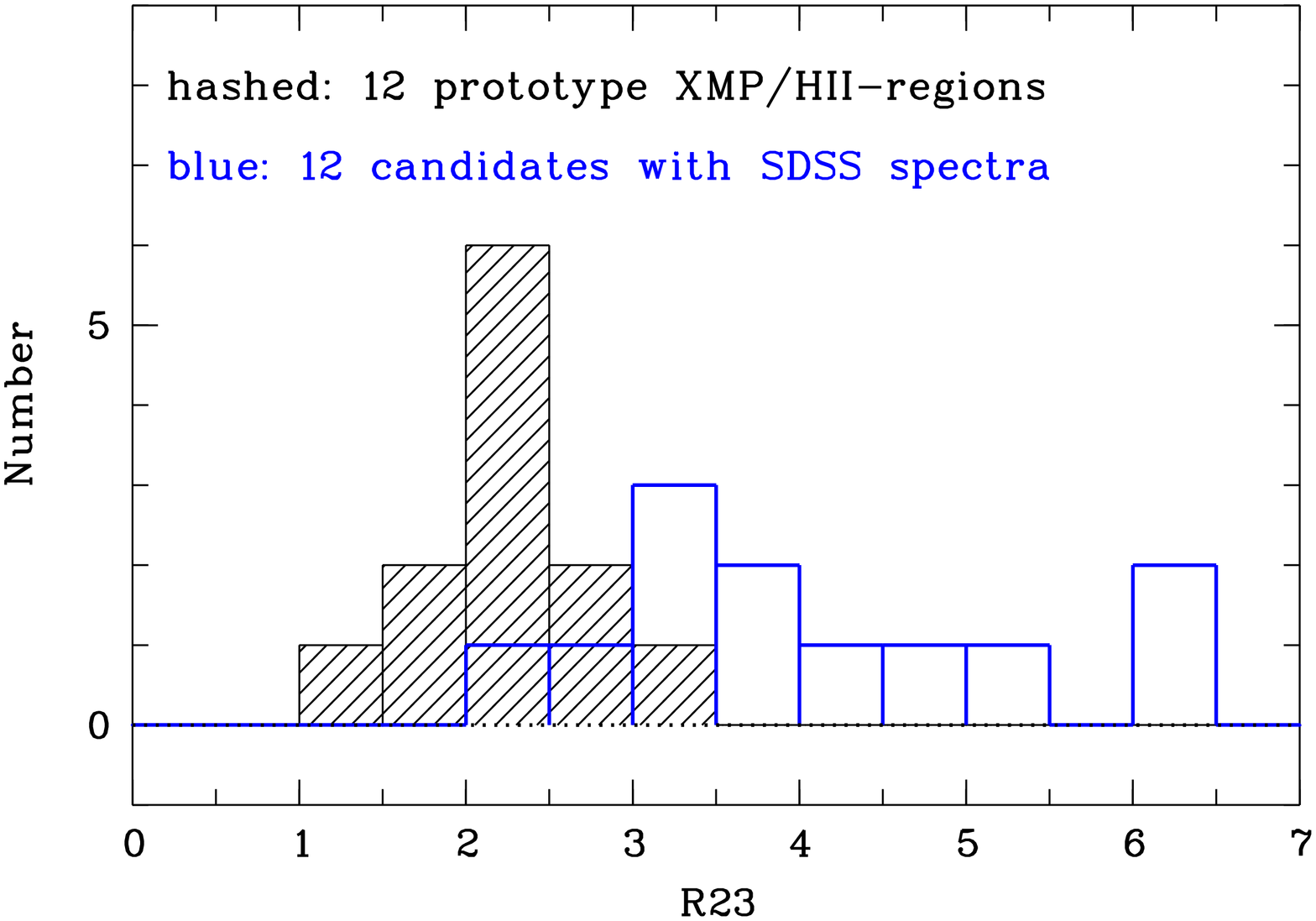}
\caption{Comparison of observational properties of 10 prototype XMP void
galaxies (hashed histograms) and XMP void candidates with available data
(blue thick histograms).
{\bf Top panel:} Distribution of the MW-extinction-corrected
integrated colour $(g-r)_{0}$. {\bf Bottom panel:} Distribution of parameter
R$_{\rm 23}$ for 12 \HII-regions in prototype XMP galaxies and 12 candidate
XMP galaxies. See text for more detail. }
\label{fig:histo2}
\end{figure}

In Figures~\ref{fig:histo1} and \ref{fig:histo2} we show distributions of
parameters  $M_{\rm B}$, MHI/$L_{\rm B}$, integrated colour $(g-r)_{0}$ and
the ratio R$_{\mathrm 23}$ for both
samples: prototype XMP void dwarfs (hashed histograms) and the sample
of selected candidate void XMP dwarfs from Table~\ref{tab:list_candidate}
(thick line histograms). As one can see, the distribution of $M_{\rm B}$ for
the candidate sample matches well that of the prototype objects. However,
$\sim$2/3 of candidates have $M_{\rm B} < -13.0$, whereas the majority
of prototype XMP objects are fainter than $M_{\rm B} = -13.0$.
 As for the ratio
MHI/$L_{\rm B}$, about 2/3 of candidates fall in the range of
MHI/$L_{\rm B}$ for prototype galaxies, but the remaining candidates have
somewhat lower gas content. An even larger difference is seen for the
distribution
of colour $(g-r)_{0}$. Only $\sim$40\% of candidates have $(g-r)_{0}$ as  
blue as for the prototype objects. As we already noticed in the description
of our selection, we decided to extend parametric space for candidate XMP in
order to account for probable errors in parameter estimates and for the
limited size of the XMP void prototype group which may affect its
representativity.

The ratio R$_{\mathrm 23}$ was estimated only for a dozen of candidates
with available SDSS spectra. Since the line
[O{\sc ii}]$\lambda$3727~\AA\ was outside the SDSS wavelength range, its
flux was estimated via the flux of [N{\sc ii}]$\lambda$6584~\AA\ as explained
above. Since the scatter in the relation between the fluxes of these two lines
can be a factor of $\sim$2, the related estimate of R$_{\mathrm 23}$ can
also have large uncertainty and should be treated only as an indicative one.
One can see that for prototype XMP dwarfs R$_{\mathrm 23}$ fall into the
range between 1.3 and 3.5. For a number of XMP candidates
R$_{\mathrm 23}$ falls in the same range. However, in general their
R$_{\mathrm 23}$ appear to be larger than for the prototype group.

\begin{figure*}
\includegraphics[width=17.0cm,angle=-0,clip=]{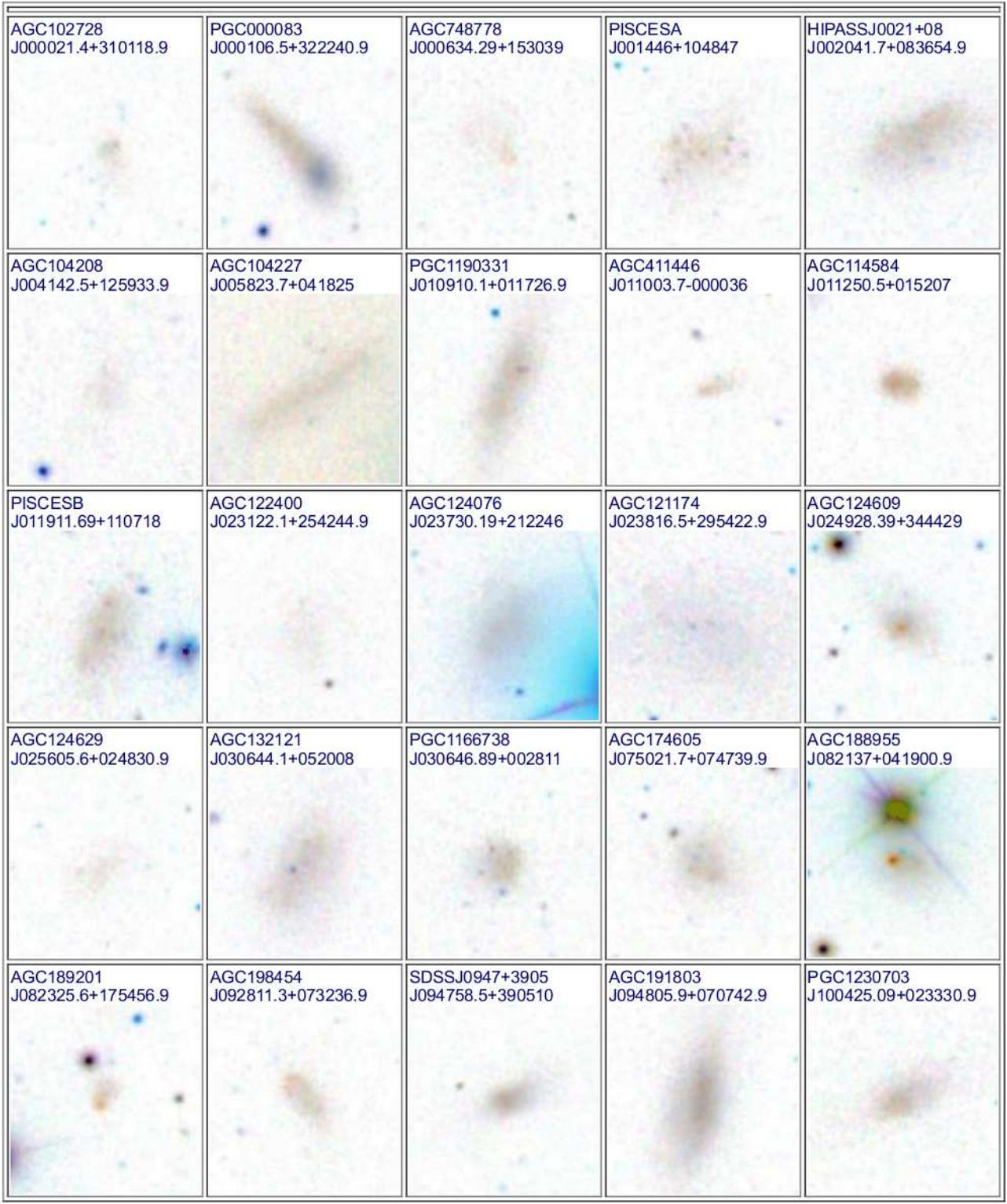}
\caption{SDSS $gri$ colour finding charts of NVG  XMP candidates falling
within the SDSS footprint.
Colours are inverted to better see very low surface brightness objects.
The first 25 galaxies are shown in RA order. The side of the box is
$\sim$50\arcsec.  North is up and east is to the left.
}
\label{fig:Cand_XMP_p1}
\end{figure*}

\begin{figure*}
\includegraphics[width=17.0cm,angle=-0,clip=]{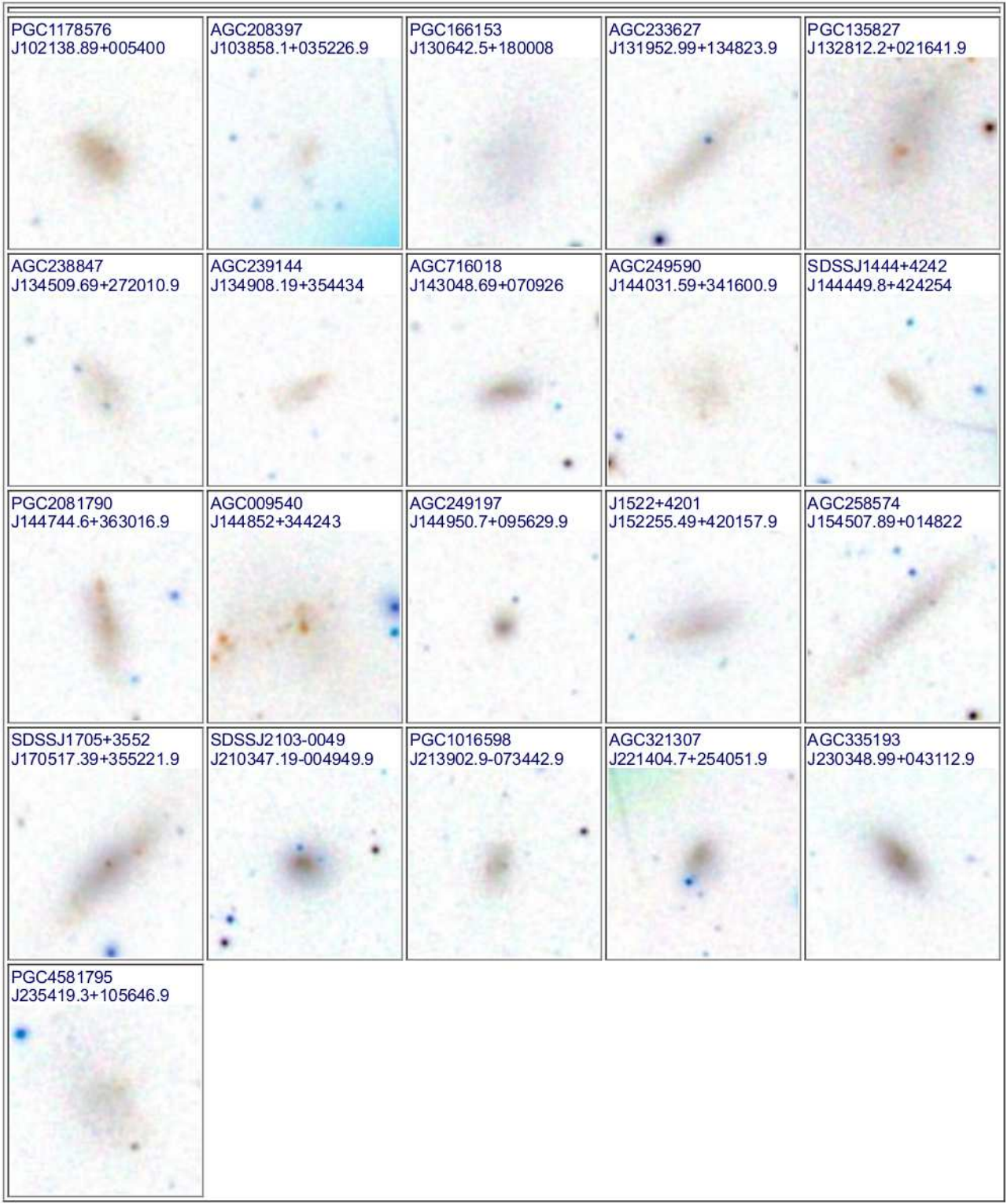}
\caption{SDSS $gri$ colour finding charts of NVG  XMP candidates falling
within the SDSS footprint.
Colours are inverted to better see very low surface brightness objects.
The last 21 galaxies are shown in RA order. The side of the box is
$\sim$50\arcsec.
}
\label{fig:Cand_XMP_p2}
\end{figure*}

\begin{figure*}
\includegraphics[width=4.0cm,angle=-0,clip=]{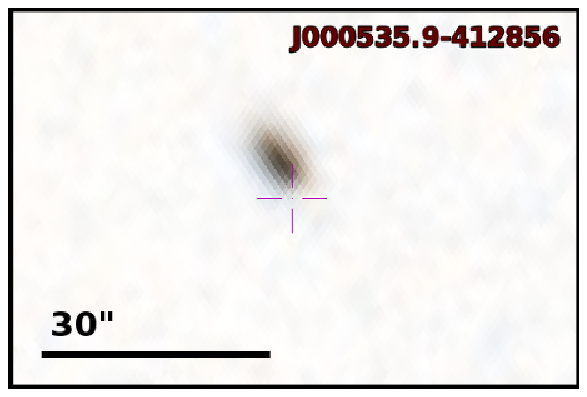}
\includegraphics[width=4.0cm,angle=-0,clip=]{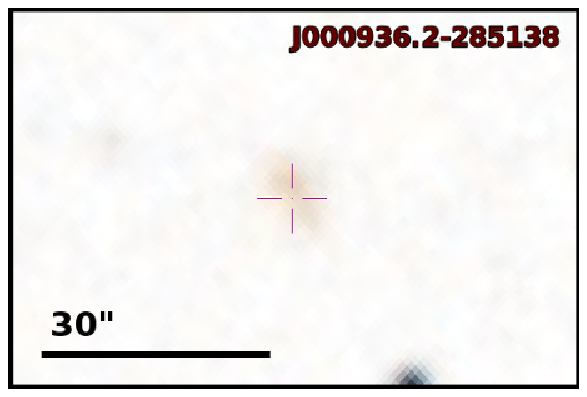}
\includegraphics[width=4.0cm,angle=-0,clip=]{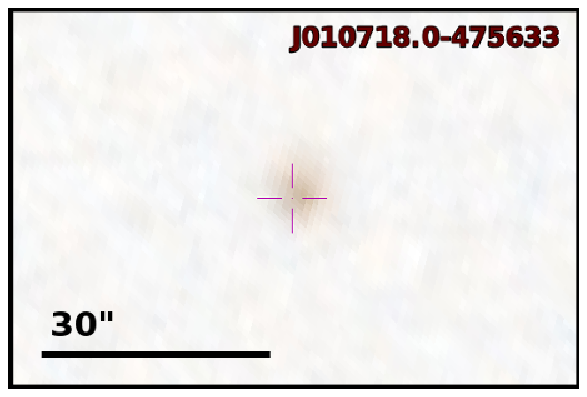}
\includegraphics[width=4.0cm,angle=-0,clip=]{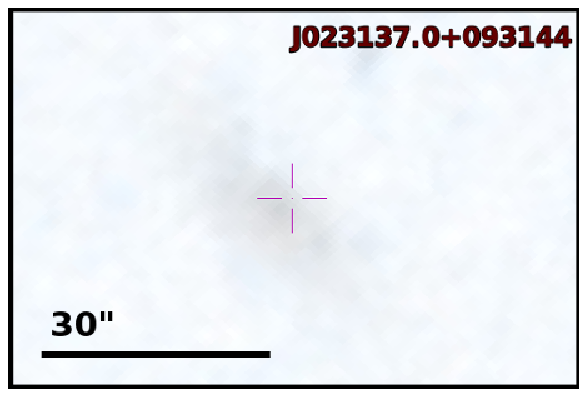}
\includegraphics[width=4.0cm,angle=-0,clip=]{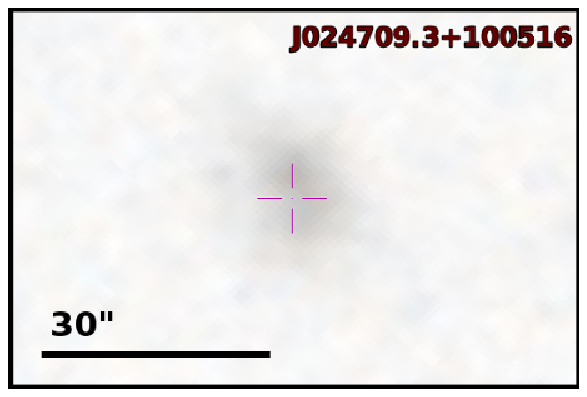}
\includegraphics[width=4.0cm,angle=-0,clip=]{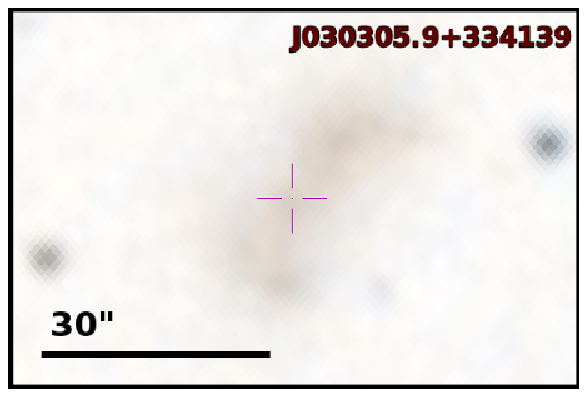}
\includegraphics[width=4.0cm,angle=-0,clip=]{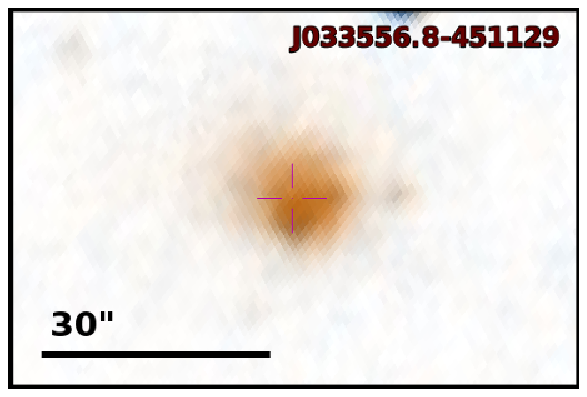}
\includegraphics[width=4.0cm,angle=-0,clip=]{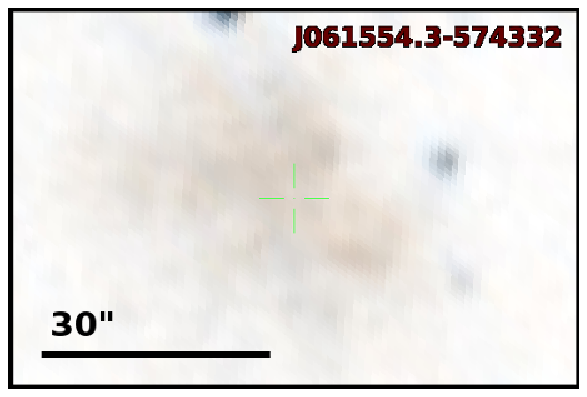}
\includegraphics[width=4.0cm,angle=-0,clip=]{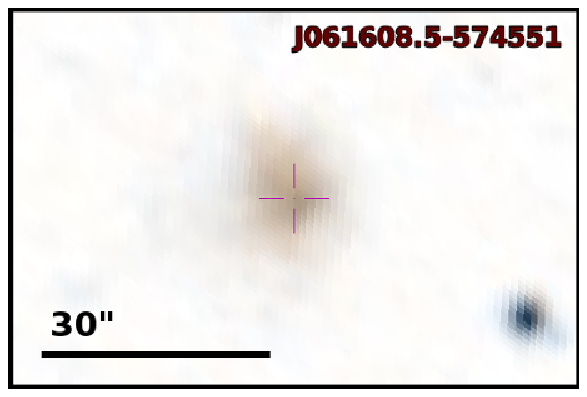}
\includegraphics[width=4.0cm,angle=-0,clip=]{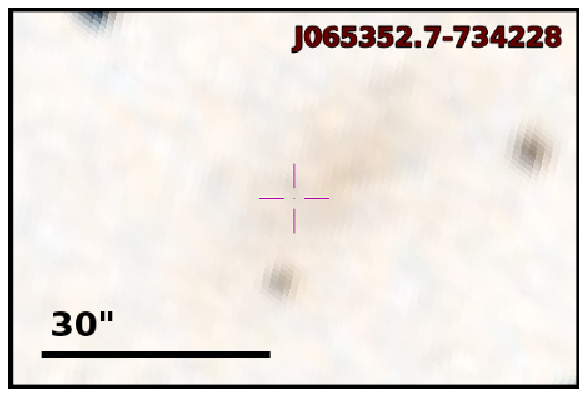}
\includegraphics[width=4.0cm,angle=-0,clip=]{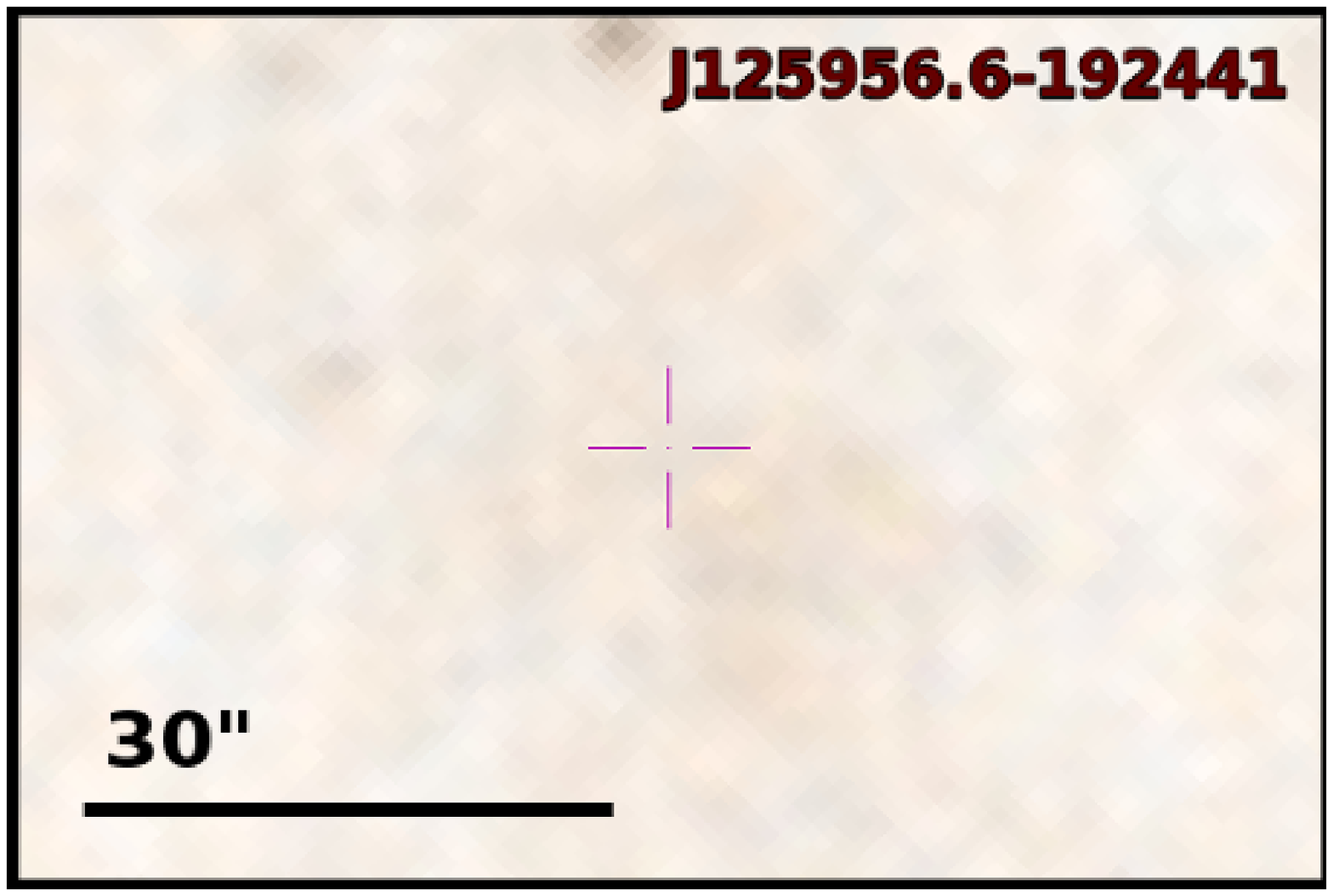}
\includegraphics[width=4.0cm,angle=-0,clip=]{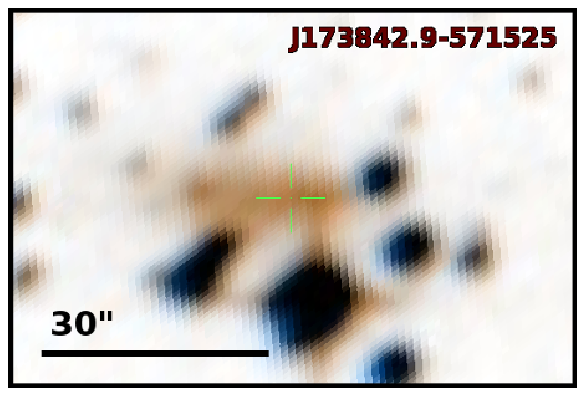}
\includegraphics[width=4.0cm,angle=-0,clip=]{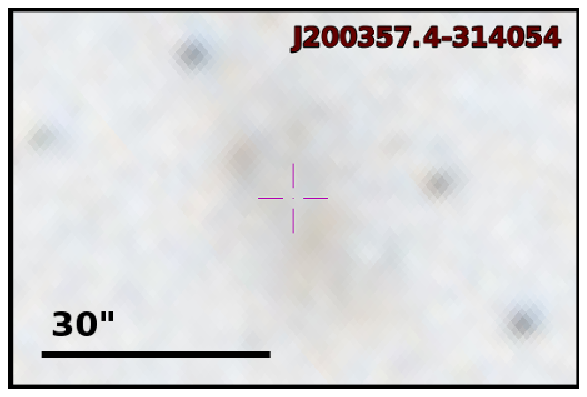}
\includegraphics[width=4.0cm,angle=-0,clip=]{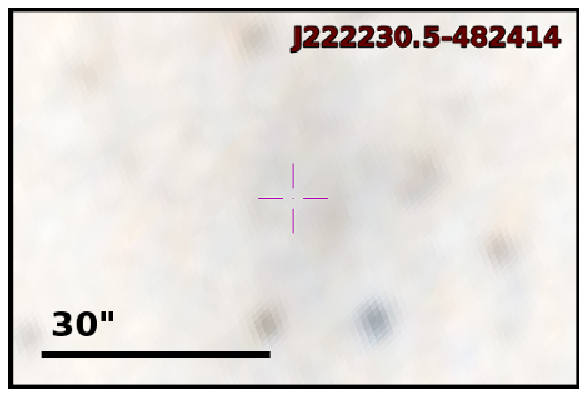}
\caption{Finding charts of 14 NVG XMP candidates falling outside the SDSS
footprint.
Colours are inverted to better see very low surface brightness objects.
Galaxies are shown in RA order. The bar in the box shows the angular scale.
}
\label{fig:Cand_XMP_nonSDSS}
\end{figure*}

In Figures~\ref{fig:Cand_XMP_p1} and \ref{fig:Cand_XMP_p2} we present
thumbnail images of 46 galaxies within the SDSS footprint. They were
obtained via the SDSS database finding chart service.
In Fig.~\ref{fig:Cand_XMP_nonSDSS} we present finding charts for 14 void
galaxies outside the SDSS zone. They are combined from the Digital Sky Survey
(DSS) $B$ and $R$ images.

\setcounter{qub}{0}
\begin{table*}
\caption{List of Nearby Void Galaxies candidate XMP dwarfs}
\begin{tabular}{r|l|l|l|r|r|r|c|l|l|p{3.5cm}}\hline \hline
\# &\MC{1}{c}{Name}&\MC{1}{c}{J2000 Coord}&V$_{\rm h}$& D &$B_{\rm tot}$&$M_{\rm B}$&$\frac{\rm MHI}{L_{\rm B}}$  &logMHI   &logM$_*$  & \MC{1}{l}{Notes} \\
   &              &                  &     & Mpc  &mag&mag&    &      &     &       \\
\MC{1}{c}{1}&\MC{1}{c}{2}&\MC{1}{c}{3}&\MC{1}{c}{4}     & \MC{1}{c}{5}&\MC{1}{c}{6}&\MC{1}{c}{7} & \MC{1}{c}{8} & \MC{1}{c}{9}  & \MC{1}{c}{10} & \MC{1}{l}{11} \\
% 1 & 2            &       3          & 4   &  5   & 6 & 7 & 8           & 9    & 10  & 11    \\
\hline
\qq&AGC102728     & J000021.4+310119 & ~566 & 9.4 &19.38&--10.65& 2.46  &6.85&5.82& \\ % Bt_new=19.38 AB=0.167 Huang+2012 logM*=5.73
\qq&PGC000083     & J000106.5+322241 & ~542 & 9.1 &17.14&--12.86& 1.41  &7.48&6.19& \\ % tadpole D~9.4 KUG 2358+320A AB=0.189
%\qq&PGC138064    & J000248.3+185807 &      &     &     &       &       &    &    & \\ % D~10.4    OUT of void sample
\qq&PGC000389     & J000535.9--412856& 1500 & 17.5&17.98&--13.30& ...   &... &... & No HI-data \\ % AB=0.059 O/H=7.77+-0.08 (se) (+Te, 4363 @1sig: 7.91+-0.09?)
\qq&AGC748778     & J000634.3+153039 & ~258 & 6.3 &18.95&--10.30& 2.30  &6.66&5.40& D(TRGB) \\ % VLSB g-r=.11 B~18.93 D~6.5 AB=0.234 Huang+2012 logM*=5.31
\qq&PGC736507     & J000936.2--285138&~898$\dagger$&...&19.14&--10.95& ...   &... & ...& not NVG object, wrong V$_{\rm h}$ \\ %  V(EL)=7594
%\qq&PGC2816192   & J001108.6+141423 &      &     &     &       &       &    &    & \\ % dI g-r=.2 XMD? B~19.3 D~11.5 OUT of void sample
\qq&PISCESA       & J001446.0+104847 & ~235 & 5.6 &17.98&--11.14& 2.92  &7.05&5.74& D(TRGB) \\ % O/H=7.31+-0.07 (se)  +KAT-7 in HI (South Africa) Tollerud+2016 logM*=7.0
\qq&HIPASSJ0021+08& J002041.7+083655 & ~693 & 9.9 &17.10&--13.37& 1.21  &7.63&6.72&  \\ % O/H=7.53+-0.08 (se)  AB=0.499
\qq&AGC104208     & J004142.5+125934 & ~676 &10.1 &19.66&--10.74& 3.45  &7.01&5.92& \\ % Image in H$\alpha$. No emission  C35 Ha SED656 (1200s)+SED606(600s) see~3.5" +GMRT (23.11.2018)
\qq&AGC104227     & J005823.7+041825 & 1198 &16.9 &18.12&--13.11& 2.13  &7.76&6.56& \\ %very faint H$\alpha$ at V(HI) no EL? M/L=2.1 Blue LSB w/blu v.faint knots FHI=0.79 W50=22 BTA 120s-image: no Ha-knots!
\qq&PGC493444     & J010718.0--475633&~837$\dagger$&...&19.47&--10.56& ...   &... &... & not NVG object, wrong V$_{\rm h}$ \\ %Wrong V(LEDA): real $z \sim$0.023. no EL? no HI? MB=-10.56  B-R=0.87
%\qq&PGC004055    & J010822.0--381233&~645  & 6.7 &16.12&--12.91& 0.57  &    &    &  \\ %  FHI=1.4  TOO low MHI/LB
\qq&PGC1190331    & J010910.1+011727 & 1094 &15.4 &17.20&--13.49& 2.24  &7.94&7.04&  \\ % O/H=7.50+-0.07 (se) =AGC111687 (+Te, 4363 @1.5sig) =PGC119033
\qq&AGC411446     & J011003.7--000036& 1137 &15.9 &19.53&--11.59& 5.14  &7.53&5.66&  \\ % O/H=7.00+-0.05 (se) DDT (~4200) +our (~2400) total ~6600s AB=0.11
\qq&AGC114584     & J011250.5+015207 & 1089 &15.4 &18.17&--13.14& 1.90  &7.62&6.02&  \\ % O/H=7.15+-0.07 (se) (+Te, 4363 @1sig: 7.16+-0.34)
\qq&PISCESB       & J011911.7+110718 & ~616 & 8.9 &17.63&--12.45& 2.58  &7.52&6.41& D(TRGB) \\ % O/H=7.40+-0.07 (se) AB=0.205  Tollerud+2016 logM*=7.5
%\qq&PGC166064    & J015520.2+275714 & ~207 & 5.7 &16.30&--12.79& 0.31  &    &    & \\ % EXCLUDE, TOO low MHI/LB
%\qq&PGC721555    & J021525.7--300518& 1499 & ..  &18.23&--13.09& 0.33  &    &    & \\ % EXCLUDE, TOO low MHI/LB
%\qq&PGC3209772   & J022349.2--285606&      &     &     &       &       &    &    & \\ % OUT of void sample ELs z=0.1274 instead of V(HyperLEDA)=540+-143!
\qq&AGC122400     & J023122.1+254245 &~938  & 15.5&18.92&--12.45& 3.66  &7.73&... & \\ % AB=0.424 bluish VLSB
\qq&AGC124137     & J023137.0+093144 &~897  & 14.0&17.98&--13.12& 2.06  &7.78&6.72& \\ % AB=0.37 XMP? SALT observed, low S/N. Need 3727
\qq&AGC124076     & J023730.2+212246 &~950  & 15.5&17.60&--13.90& 2.51  &8.15&... & \\ % AB=0.545
\qq&AGC121174     & J023816.5+295423 &~693  & 12.4&16.89&--14.12& 6.55  &8.39&7.65& \\ % AB=0.542 KKH013 LSB g-r=.55  CHECK MB ! FHI=6.75
\qq&AGC123223     & J024709.3+100516 &~767  & 12.4&17.52&--13.77& 1.79  &7.99&7.01& \\ % AB=0.80 O/H=7.49+-0.07 (se) A_B=0.96?
\qq&AGC124609     & J024928.4+344429 &1588  & 25.0&18.00&--14.25& 1.31  &8.05&7.07& \\ % AB=0.266 LSB+knot g-r=.14 B~18.2 AB=0.32 MB~-14.1
\qq&AGC124629     & J025605.6+024831 &~794  & 12.4&19.46&--11.50& 1.72  &7.22&5.88& \\ % AB=0.397  O/H=7.00+-0.05 (se) I3727 is ~1.3IHb, VLSB, NEED more!
\qq&KKH18         & J030305.9+334139 &~210  &  4.8&17.11&--12.01& 1.90  &7.20&6.49& D(TRGB) \\ % PS1 photo; NII/H$\alpha \sim$0.03$\pm$0.02. SII/H$\alpha \sim$0.17  1800R ~3000s at var.transp. +SED656 120s. Karach+2014: logM*=7.21 from K-luminosity
%--------------------
\qq&AGC132121     & J030644.1+052008 &~678  & 11.0&17.30&--13.53& 1.62  &7.86&6.79& \\ % AB=0.624 O/H=7.40+-0.07 (se)  W50=43
\qq&PGC1166738    & J030646.9+002811 &~710  & 11.2&17.94&--12.67& 0.89  &7.21&6.39& \\ % AB=0.377  ?? SDSS weak emiss (Ha,5007). may be XMP van Driel+2016 logM*=6.82
\qq&PGC013294     & J033556.8--451129&~737  &  7.3&16.35&--13.36& ...   &... &6.82& D(TRGB) No HI-data \\ % O/H_Te=7.65 blue dIrr +UV 6dFJ0335568-451129, FHI < 1 Jy*km/s (???)  Karach+14 logM*=7.62
\qq&ESO121-020    & J061554.3--574332&~582  &  6.1&15.87&--13.20& 3.87  &8.06&6.60& D(TRGB) \\  % des. Karach+2014 logM*=7.07  AB=0.151 E+A EW_abs~3A faint 5007/Hb~0.5: XMP! NEED Ha image
\qq&PGC385975     & J061608.5--574551&~554  &  6.1&17.10&--11.79& 2.49  &7.38&6.03& comp. of ESO121-020 \\ % des. Karach+2014 logM*=7.62 NEED new AB!  O/H=7.46+-0.07 (se) On CCD center
%\qq&PGC888024    & J061753.8--170905&~850  & ... &...  &--13.35& ...   &    &    & HIPASS       \\ %  No in Catalog final version!
\qq&HIPASSJ0653-73& J065352.7--734228&1208  & 13.5&17.90&--13.14& 5.39  &8.19&8.03&  \\ %  AB=0.484   logM* (Gurovich+2010)
%\qq&DDO47        & J074155.4+164809 &~272  &  514&13.60&--15.77&       &    &    & \\ % AB=0.12 M/LB~2.7  TOO BRIGHT?
\qq&AGC174605     & J075021.7+074740 &~351  &  9.9&17.82&--12.25& 1.33  &7.21&6.44& D(TRGB) \\ % logM*=6.65 (Karach.+2014) Only Ha-emis Vh=351 D(TRGB)=9.9 B~18.7 AB=0.084  Ha-image in Hauberg+2015
\qq&AGC188955     & J082137.0+041901 &~758  & 12.8&17.57&--13.04& 0.86  &7.40&... & very blue \\ % not gas-rich? AB=0.078 In SF knot broad EL wings O/H~7.65? broad component?
\qq&AGC189201     & J082325.6+175457 &1475  & 23.4&19.24&--12.74& 4.03  &7.89&6.09& \\ % AB=0.136  O/H$\sim$7.1:  1800R 1800s NII/Ha <0.02 SII/Ha~0.06 see.~3" + GMRT (25.01.2019)
\qq&AGC198454     & J092811.3+073237 &1373  & 21.0&18.51&--13.28& 1.25  &7.84&6.54& \\ % AB=0.183 O/H~7.50+-0.08 (se) pg900 strong lines +4363? B_new~18.52
\qq&SDSSJ0947+3905& J094758.5+390510 &1501  & 24.9&18.03&--14.01& 3.20  &8.17&7.22& in triplet with MRK407  \\ %  AB=0.06 FHI has substant. uncert. Need GMRT map estimate
\qq&AGC191803     & J094805.9+070743 &~526  &  9.2&16.79&--13.14& 1.38  &7.58&7.01& \\ % AB=0.096  O/H=7.48+-0.09 (se) NII/Ha~0.05+-0.05 SDSS XMP-cand
%\qq&J1001+0846   & J100109.5+084656 &1265  & 19.2&18.09&--13.57& 0.30  &    &    & \\ % O/H=7.62+-0.07 (se) SDSS, NIBLES0976 FHI=0.13  was REMOVED as possb. Leo Cloud member
\qq&PGC1230703    & J100425.1+023331 &1126  & 17.1&17.71&--13.53& 0.56  &7.16&7.04& \\ % logM*=7.12 (Butcher+2016) O/H=7.68+-0.07 (se) NIBLES0987 FHI=0.33
\qq&PGC1178576    & J102138.9+005400 &~701  & 11.0&17.35&--13.04& 1.44  &7.58&6.40& \\ % logM*=7.01 (Karach+2014) AB=0.187 O/H=7.36+-0.07 (se) MB=-13.15 MHI/LB=1.3 SDSS NII/Ha<0.015 SALT NII/Ha~0.02+-0.02
\qq&AGC208397     & J103858.1+035227 &~763  & 11.9&19.95&--10.56& 5.75  &7.18&5.60& \\ % AB=0.141  NII/H$\alpha \sim$0.03  1800R 2700s NII/Ha=0.03+-0.003    +GMRT (02.02.2019)
\qq&PGC044681     & J125956.6--192441&~827  &  7.3&17.50&--12.16& 4.88  &7.77&6.64& D(TRGB) \\ % PS1; logM*=7.39 (Karach+2014) AB=0.352  O/H=7.38+-0.10 (se) OIII5007~0.28*Hb, OII3727~1.9*Hb, Is SE-method applicable?
\qq&PGC166153     & J130642.5+180008 &1573  & 22.5&18.03&--13.82& 5.11  &8.42&7.77& \\ %  VLSB  AB=0.092 NED: Vh=-178? (SDSS)
\qq&AGC233627     & J131953.0+134824 &~937  & 13.4&17.78&--12.94& 2.76  &7.81&6.87& \\ %  LSB   AB=0.089
\qq&PGC135827     & J132812.2+021642 &1008  & 13.5&16.51&--14.24& 1.62  &8.09&7.20& \\ % AB=0.095  O/H_k1=7.74+-0.11 (Te) NII/Ha~0.02+-0.004 O/H_k2=7.87+-0.13 NII/Ha~0.04+-0.004
%\qq&AGC233202    & J133133.9+021118 &1136  & ..  &17.8 &--13.19& 1.40  &    &    & PGC135829    \\ % Neither in final sample, nor in initial!
\qq&AGC238847     & J134509.7+272011 &~901  & 13.5&18.57&--12.21& 2.64  &7.47&6.40& \\ % AB=0.045
\qq&AGC239144     & J134908.2+354434 &1366  & 20.4&19.06&--12.54& 3.17  &7.70&6.01& \\ % O/H=6.95+-0.10 (se) 3600s see=1.2" +GMRT (25/26.01.2019)
\qq&AGC716018     & J143048.7+070926 &1365  & 19.1&18.18&--13.19& 3.72  &8.08&6.98& \\ % logM*=7.04 (van Driel+2016)  AB=0.093
\qq&AGC249590     & J144031.6+341601 &1489  & 21.7&18.45&--13.27& 2.70  &7.93&6.62& \\ % AB=0.044      +GMRT (21.01.2019)
\qq&SDSSJ1444+4242& J144449.8+424254 &~634  & 10.9&19.11&--10.54& ...   &... &5.71& near UGC9497, no HI-data  \\ % O/H=7.17+-0.07 (se) DDT BTA 80min @seeing=2.4" +GMRT (5.03.2018) MHI/LB~0.75 lgMHI=6.53
\qq&PGC2081790    & J144744.6+363017 &1226  & 18.1&17.66&--13.67& 1.49  &7.83&6.95& \\ % logM*=6.62 (van Driel+2016) O/H=7.45+-0.08 (se) 3600s low S/N 4363 90min  NIBLES
\qq&AGC009540     & J144852.0+344243 &~801  & 12.1&16.25&--14.27& 2.29  &8.26&7.09& \\ %
\qq&AGC249197     & J144950.7+095630 &1809  & 24.2&18.69&--13.48& 2.45  &7.92&6.75& \\ % AB=0.101 BCG  CHECK! No our photometry?
\qq&J1522+4201    & J152255.5+420158 &~608  &  9.8&17.87&--12.18& ...   &...&6.63& No HI-data, SDSS spect \\ % MHI/LB=0.49 (GMRT) O/H=7.30+-0.10 (se) 1800s @seeing=1.5" +GMRT (20.03.2018) lgMHI=6.75
\qq&AGC258574     & J154507.9+014822 &1523  & 19.3&17.72&--13.09& 2.60  &8.20&7.18& \\ % O/H_k2=7.32+-0.12 (se) NII/Ha~0.05+-0.035 O/H_k1=7.60+-0.05 NII/Ha~0.06+-0.04
\qq&SDSSJ1705+3552& J170517.4+355222 &~992  & 14.1&16.99&--13.85& ...   &... &7.20& No HI-data, SDSS spect \\ % AB=0.085  SDSS spec. XMD?
\qq&HIPASSJ1738-57& J173842.9--571525&~858  &  7.3&16.62&--13.06& 2.18  &7.73&... &  \\ % AB=0.351 FHI=4.49  O/H(se)mean,c~7.56
\qq&KK246         & J200357.4--314054&~431  &  7.1&17.07&--13.28& 2.30  &8.02&7.70& D(TRGB) \\ % logM* (Kirbi+2008)   AB=1.10 O/H=7.69+-0.07 (se) NHII/Ha~0.04+-0.01 CHb=0.6 Large diff. w/Nicholls+2014 (O/H=8.17) Their low A_V~0.035 is very suspec., since A_B(MW)=1.34!
\qq&SDSSJ2103-0049& J210347.2--004950&1411  & 17.4&17.44&--14.02& 1.36  &7.83&7.33& \\ % AB=0.258 O/H=7.31+-0.07 (se)  75min @seeing=1.65"
\qq&PGC1016598    & J213902.9--073443&1283  & 15.4&18.63&--12.43& ...   &... &6.63& No HI-data \\ % AB=0.142 O/H=7.30(s)
%\qq&AGC322279    & J220316.8+174744 &      &     &     &       &       &    &    & \\ % 1800R  OUT of voids
\qq&AGC321307     & J221404.7+254052 & 1152 & 16.2&18.29&--13.07& 2.85  &7.86&7.01& \\ % AB=0.307  O/H=7.78+-0.08 (se)  and 20170913: NII/Ha<0.03
\qq&ESO238-005    & J222230.5--482414& ~706 &  8.0&15.53&--13.59& 2.16  &8.34&8.19& D(TRGB) \\ % O/H=7.33 (s) logM* (Karach+2014) W50~80 FHI=16.3 (EDD) 2-horn prof. KK257 O/H(s)-0.03=7.33
\qq&AGC335193     & J230349.0+043113 & 1125 & 16.1&17.25&--14.03& 0.56  &7.58&7.25& \\ % AB=0.236  O/H=7.60+-0.07 (se) MB=-14.2
%\qq&AGC332939    & J230816.7+315406 &      &     &     &       &       &    &    & \\ % OUT of voids O/H=7.70+-0.07 (se) and 20170913: NII/Ha~0.015
\qq&PGC4581795    & J235419.3+105647 & ~943 & 13.2&17.52&--13.41& 1.04  &7.53&6.97& AGC332123=F750-V01 \\ %  Only Ha at V~900 at SALT FHI=0.9 AB=0.328
\hline
\multicolumn{11}{p{16.2cm}}{Candidate galaxy parameters are presented as follows: Col.~1: $id$ number; Col.~2: common name; Col.~3: J2000. coordinates;} \\
\multicolumn{11}{p{16.2cm}}{Col.~4: radial velocity in \kms; Col.~5: adopted distance (see Table description); Col.~6: Total $B$ magnitude; Col.~7: absolute } \\
\multicolumn{11}{p{16.2cm}}{$B$ magnitude; Col.~8: MHI/$L_{\rm B}$ ratio in solar units; Col.~9: logarithm of MHI\ in M\sunn, Col.~10: logarithm of M$_{*}$ in M\sunn.}\\
\multicolumn{11}{p{16.2cm}}{$\dagger$ galaxies marked as 'wrong V$_{\rm h}$' appear distant based on SALT data.} \\
\end{tabular}
\label{tab:list_candidate}
\end{table*}
% J0000+3101 M* Huang+2012 (2012AJ....143..133) SED
% J0006+1530 M* Huang+2012 (2012AJ....143..133) SED
% PISCESA lgM*=7.0 Tollerud+2016 (2016ApJ...827...89T) vs our 5.74
% PISCESB lgM*=7.5 Tollerud+2016 (2016ApJ...827...89T) vs our 6.41
% KKH18=J0303+3341 M* Karachentsev+2014 (2014AJ....147...13K)
% J0306+0028 lgM*=6.82 van Driel+2016 (2016A&A...595A.118V) vs our 5.88
% J0335-4511 M* Karachentsev+2014 (2014AJ....147...13K)
% ESO121-020=J0615-5743 lgM*=7.78 from Karachentsev+2014 (2014AJ....147...13K) vs our 6.60
% J061608.5-5745 lgM*=7.07 Karachentsev+2014 (2014AJ....147...13K) vs our 6.03
% J0653-7342  lgM*=8.03 Gurovich+2010 (2010AJ....140..663G)
% J0750+0747  lgM*=6.65 Karachentsev+2014 (2014AJ....147...13K)
% J1004+0233 lgM*=7.12 Butcher+2016 (2016A&A...596A..60B)
% J1021+0054 lgM*=7.01 Karachentsev+2014 (2014AJ....147...13K)
% J1259-1924 lgM*=7.39 Karachentsev+2014 (2014AJ....147...13K) vs our - 6.64
% J1430+0709 lgM*=7.04 van Driel+2016 (2016A&A...595A.118V)
% J1447+3630 lgM*=6.62 van Driel+2016 (2016A&A...595A.118V)
% KK246=J2003-3140 lgM*=8.15  Shao+2018 (2018MNRAS.479.3509S)
% J2222-4824 lgM*=8.19 Karachentsev+2014 (2014AJ....147...13K)

\section[]{Discussion}
\label{sec:discussion}

\subsection[]{Scenarios for various types of XMP dwarfs}
\label{ssec:scenarios}

It is worth underlining that XMP dwarfs can have rather different
evolutionary histories. Four physical mechanisms resulting in the
observed very low gas metallicity in galaxies have been proposed and have
observational support. Two of them  define the secular
galaxy evolution. These are A) a less efficient star-formation in galaxies
of smaller mass and the related slower metal production (so-called
'Downsizing') and B) the increased metal loss via the galactic wind
in smaller galaxies due to their shallow galaxy gravitational well
\citep[e.g.,][]{Garnett02}. One can also suggest the interplay between
the two mechanisms.

In periods of elevated
star formation the collective effect of multiple SNe leads to galaxy winds
which efficiently evacuate the fresh metals into the intergalactic medium.
As shown by \citet{McQuinn15}, the nearby very low mass XMP dwarf
Leo~P, with 12+$\log$(O/H)=7.17~dex has lost $\sim$95\% of its produced metals
during its long life. This mechanism is quite general and seems to
contribute substantially to
the known direct relation between metallicity and galaxy mass
or luminosity. One such tight  relation is exemplified by \citet{Berg12}
using a well defined sample of late-type galaxies in the Local Volume
with the $M_{\rm B}$ range of --10.5 to --18.5. It extends to even
lower $M_{\rm B}$, since the very low value of O/H for Leo~P with
$M_{\rm B}$ = --8.8 matches the relation of 12+$\log$(O/H) versus
$M_{\rm B}$ from \citet{Berg12} well.

Two other mechanisms, leading to significantly reduced gas metallicity,
sometimes falling into the XMP regime, are transient and
more localized, and thus can be considered as additional factors
affecting the secular evolution of gas metallicity.
The characteristic time-scale for erasing the temporary O/H drop
is comparable to the gas disc rotational period. The third scenario
is expected for the SF bursts in the periphery of galaxy discs induced
by accretion events of very metal-poor ambient intergalactic gas with
Z $\sim$(0.01--0.02)~Z\sunn\ (the so called 'Cold Accretion' along 
cosmic web filaments, e.g., \citet{Sanchez16,Ceverino16}). Due to the
admixture of the fresh ambient gas, the observed metallicity in a SF
region can be substantially reduced with respect to the global value
for a galaxy in question.

A similar mechanism is related to the transient event of a
(central) starburst induced by interaction with a galaxy
disturber \citep{Ekta2010}.
As shown by e.g. \citet{Thuan2005}, the distant periphery
of discs in late-type dwarf galaxies can keep metallicity at the
levels of the ambient intergalactic medium. When this gas is
substantially disturbed, it can lose stability and inflow into
the galaxy center with the subsequent mixture and an induced starburst.
Thus, this phenomenon can substantially dilute the metallicity
of the induced SF burst and temporarily appear as a
very metal-poor object.

As follows from information in Col. 12 of Table~\ref{tab:prop_summary},
a substantial fraction of prototype XMP void dwarfs appear to be in pairs
or triplets.
Here we provide the references supporting this evidence. J0113+0052 is
suggested as a minor merger due to the detached H$\alpha$ kinematics in
its southern component \citep{Moiseev2010} and highly disturbed HI
morphology and kinematics \citep{Ekta2008}. J0706+3030 is very likely
a separate faint component in a merging triplet, UGC3672 \citep{U3672}.
J0943+3326 (AGC198691), with a radial velocity difference of only 37~\kms\
and a projected distance of $\sim$40~kpc at the adopted D=10.8~Mpc
\citep{Hirschauer16,PaperVII}, has clearly disturbed HI gas and
is very likely a fainter companion of a nearby more massive dwarf, UGC5186.
J0956+2850 (DDO68B), an outer part of the well-known XMP galaxy, DDO68
\citep{DDO68,Ekta2008,Annibali2016,DDO68_HST}, is treated as a detectable
remnant of a recent minor merger.
J2104--0035, despite being well isolated, appears disturbed in HI
GMRT maps. This led \citet{Ekta2008} to suggest that it could be an advanced
merger. Finally, two extremely gas-rich dwarfs with unknown gas metallicity,
J0723+3622
and J0723+3624, are smaller members of a triplet, J0723+36, near a void center
\citep{Triplet}.

Therefore, in general, the latter two mechanisms are expected to
contribute to the observed low metallicity of void XMP dwarfs.
However, in voids, significantly reduced gas metallicity can occur
due to the specific environment and related conditions
of galaxy formation and evolution. They include the late formation
of low-mass dark matter halos and the reduced star formation rate
due to rare galaxy interactions. Thus, the very fact that a galaxy
resides in a void, increases the possibility of its discovery as a
less evolved object. In addition, the abovementioned processes of
transient metallicity reduction can increase the probability
of finding void XMP dwarfs.

\subsection[]{Very Young Galaxies}
\label{ssec:VYG}

As formulated in the Introduction, the main goal of this project is
the search for new unusual void dwarfs resembling Very Young Galaxies
in their properties. These are expected to exist in the Local Universe
according to simulations of \citet{Tweed18}.
They are defined as objects that formed more than half their stellar
mass during the last $\sim$1 Gyr.

\citet{Izotov18,Izotov19a} show from examples of the two record-low XMP
star-forming galaxies that the best candidates for objects without traces
of old stellar populations appear to be the lowest metallicity dwarfs.
This is consistent with the idea that their extremely low metallicity is
related to
the short time elapsed since the beginning of their main SF episode
and the related low metal enrichment of their gas.

A similar conclusion was drawn in our study of
dwarfs in the Lynx-Cancer void \citep{PaperIV,PaperVII}.
Namely, we found that some of the most metal-poor void dwarfs, with
gas O/H approaching 1/50--1/30 of the solar value, show unusually blue $ugri$
colours in the outer parts, outside the regions of current SF. In contrast to
the majority of other void galaxies, in two-colour diagrams they appear
close to the positions corresponding to the
{\it PEGASE} evolutionary tracks for ages of star-formation episodes
of $\lesssim$1--3~Gyr.
In the accompanying paper (in preparation), we explore
colours of the underlying LSB discs of newly discovered void XMP dwarfs.
 Based on the available SDSS images, we show that several of them have
similar $ugri$ colours corresponding to relatively small ages of
visible stellar population. However, due to their LSB nature and the faint
magnitudes, getting  deeper images of these XMP objects and
more accurate colours will be crucial for a more convincing conclusion.

On the other hand, as our previous study shows \citep{PaperVII},
voids provide a conducive environment for slower galaxy evolution
and related reduced gas metallicity in comparison to similar objects
in denser environments.
Recent models \citep{Einasto11} and N-body simulations with higher
mass resolution \citep{Rieder13} also indicate the long timescale
effect of void environments. In particular, they predict a
significant lag in the epoch of formation of
lower mass DM haloes in voids with respect to
similar epochs in denser environments.
The expected delayed galaxy formation in voids is the natural
cause of the observed reduced gas metallicity of void galaxies
in addition to the four mechanisms discussed above.

\section[]{Summary and Conclusions}
\label{sec:conclusions}

The search for the most metal-poor dwarfs has been a popular direction in
extragalactic research over almost half a century. In connection with
the recent model predictions of rare Very Young Galaxies, this work is
even more relevant since it allows the observed number of such objects
to be related to the expected numbers in different variants of possible
cosmological models.

As discussed in Sect.~\ref{sec:XMP_prototype}
and Sect.~\ref{ssec:VYG}, some of the prototype void XMP dwarfs, as well as
IZw18C, have properties that resemble those of VYGs predicted in simulations.
Looking ahead, we state that several of our new void XMP galaxies
presented in the accompanying papers, are very similar in their properties
to the prototype XMP galaxies.
Therefore, systematic methods for finding gas-rich XMP dwarfs as the best
proxies for VYGs, can provide us with an opportunity to
estimate a lower limit for the real fraction of VYGs in
the nearby Universe, and in this context are potentially
promising as one of the observational approaches to the VYG issue.

Summarizing the presented results and related discussion, we draw the
following conclusions:

\begin{enumerate}
\item
The NVG sample provides us with a new opportunity to search for unusual
XMP gas-rich dwarfs in voids, similar to known prototype XMP dwarfs.
\item
For a more efficient search for new XMP dwarfs among objects in the
NVG sample, we summarize the observational properties of 8 known void XMP
dwarfs (prototype objects) and 2 similar very gas-rich dwarfs without O/H
and use them to separate the most probable void XMP candidates.
Most of the necessary data for the selected candidates are obtained from
the public databases HyperLEDA, NED, SDSS and from the literature,
and when possible, have been checked and updated by the authors.
\item
The main selection criteria, as outlined in Sec.~\ref{sec:XMP_candidates},
are the similarity in properties to the prototype XMP dwarfs, namely,
the residence in Nearby Voids, the low luminosity, the elevated HI-content
and the blue/bluish colours.
The available information on galaxy spectra with an indication of
possible very low O/H was used as well as an additional criterion. As a
result, we identify
sixty void XMP candidates within the $M_{\rm B}$ range of [--10.3, --14.3]
for a subsequent more careful study.
\item
The suggested selection method has proved to be very efficient. This is
supported by the results of the spectroscopy for half of the selected XMP
candidates presented in two accompanying papers (Pustilnik et al.,
MNRAS, submitted).
A dozen new void dwarfs with 12+$\log$(O/H) $\lesssim$ 7.3~dex were found.
Six of them have 12+$\log$(O/H) in the range of $\sim$7.0--7.17~dex.
Their properties are very similar to those of void gas-rich XMP
prototypes.
The additional checks and constraints of  the mass of old stellar
population for this group should provide the better understanding of their
origin and evolution. At the moment, the latter objects seem  to be the
best known nearby proxies of predicted in models Very Young Galaxies.
\item
Since the original NVG sample is a volume limited one and includes
all known galaxies, one can estimate the selection function and obtain 
realistic estimates of the sample completeness for various stellar masses.
However, to further exploit the potential of these very gas-rich
XMP dwarfs for the issue of VYGs, we need in some steps to confirm their
direct relation to the simulated VYGs.
Hopefully, this will allow for a more realistic estimate of a lower
limit of the number of potential VYG objects in various ranges of galaxy
stellar or
baryonic mass and to make a preliminary comparison with model and
simulation predictions.
\item
The general conclusion of the efficiency of search for VYG proxies in
nearby voids should stimulate modelers to advance simulations in order
to include in their results a more detailed characterization of the VYG
surroundings.

\end{enumerate}

\section*{Acknowledgements}
The authors acknowledge the support from Russian Foundation for Basic Research
(RFBR) grant No.~18--52--45008. AYK acknowledges support from the
National Research Foundation of South Africa.
The authors thank the referee J.~Sanchez~Almeida for helpful comments and
stimulating questions which allowed us to improve the paper presentation.
We thank J.~Menzies for general check and improvement of the paper language.
The usage of the HyperLEDA database is greatly acknowledged.
This research has made use of the NASA/IPAC Extragalactic Database (NED)
which is operated by the Jet Propulsion Laboratory, California Institute
of Technology, under contract with the National Aeronautics and Space
Administration.
We acknowledge the use of the SDSS database.
Funding for the Sloan Digital Sky Survey (SDSS) has been provided by the
Alfred P. Sloan Foundation, the Participating Institutions, the National
Aeronautics and Space Administration, the National Science Foundation,
the U.S. Department of Energy, the Japanese Monbukagakusho, and the Max
Planck Society. The SDSS Web site is http://www.sdss.org/.
The SDSS is managed by the Astrophysical Research Consortium (ARC) for the
Participating Institutions.
We acknowledge the use for this work of public archival data from the
Dark Energy Survey (DES) and the PanSTARRS1 Surveys (PS1) and the PS1
public science archive.
%===========================================================================

\bsp

\label{lastpage}

\end{document}